\documentclass[twocolumn,superscriptaddress,article]{revtex4-2}
\usepackage{overpic}
\usepackage{graphicx}
\usepackage{graphics}
\usepackage{amssymb, amsmath}
\usepackage{color}
\usepackage{multirow}
\usepackage{epstopdf}
\usepackage{mdframed}
\usepackage{rotating}
\usepackage{tikz}
\usepackage{pgfplots}
\usepackage{standalone}
\pgfplotsset{compat=newest}
\newcommand{\ud}{\mathrm{d}}
\newcommand{\mr}{\mathbf{r}}

\newcommand{\mE}{\mathbf{E}}
\newcommand{\mH}{\mathbf{H}}

\newcommand{\mJ}{\mathbf{J}}
\newcommand{\me}{\mathbf{e}}

\newmdtheoremenv{defin}{Definition}

\begin{document}

\title{High-accuracy Casimir-Polder force calculations\\ using the Discontinuous Galerkin Time-Domain method}
\author{Philip Tr\o st Kristensen}
\affiliation{DTU Electro, Technical University of Denmark, {\O}rsteds Plads 343, 2800 Kgs.~Lyngby, Denmark}
\affiliation{NanoPhoton - Center for Nanophotonics, Technical University of Denmark, {\O}rsteds Plads 345A, 2800 Kgs.~Lyngby, Denmark}
\author{Bettina Beverungen}
\author{Francesco Intravaia}
\affiliation{Institut f\"ur Physik, Humboldt Universit\"at zu Berlin, Newtonstra{\ss}e 15, 12489 Berlin, Germany}
\author{Kurt Busch}
\affiliation{Institut f\"ur Physik, Humboldt Universit\"at zu Berlin, Newtonstra{\ss}e 15, 12489 Berlin, Germany}
\affiliation{Max-Born-Institut, Max-Born-Stra{\ss}e 2a, 12489 Berlin, Germany}

\date{\today}

\begin{abstract}
We describe a numerical time-domain approach for high-accuracy calculations of Casimir-Polder forces near micro-structured materials. 
The use of a time-domain formulation enables the investigation of a broad range of materials described by advanced material models, including nonlocal response functions. We validate the method by a number of example calculations for which we thoroughly investigate the convergence properties of the method, and comparing to analytical reference calculations, we find average relative errors as low as a few parts in a million. As an application example, we investigate the anisotropy-induced repulsive behavior of the Casimir-Polder force near a sharp gold wedge described by a hydrodynamic Drude model. 
\end{abstract}

\maketitle

\section{Introduction}
\vspace{-1.5mm}
The Casimir-Polder force~\cite{Casimir_PhysRev_73_360_1948} is a force that acts on polarizable particles due to scattering of quantum and thermal electromagnetic fluctuations in the environment. In its nonretarded limit, where it is known as the van der Waals force, it was originally investigated by London to describe the interaction between closely spaced neutral atoms or molecules. 
The theory was subsequently extended to larger separations by Casimir and Polder by including the effects of retardation. Along with its associated potential, it plays an important role in several areas of science ranging from chemistry to physics. In modern quantum technologies, such as atom-chips \cite{Henkel99, Reichel11, Schneeweiss12, Keil16, Wongcharoenbhorn21}, atom-interferometers \cite{Cronin09,Hornberger12, Alauze18}, and atom-fiber systems \cite{Vetsch10, Reitz13}, 
the Casimir-Polder force represents both a challenge and a useful tool in the design of suitable trapping potentials. From a broader perspective, the Casimir-Polder force effectively provides the leading order approximation to the Casimir force between macroscopic bodies, and it therefore serves as a convenient tool for estimating the qualitative behavior of the latter for sufficiently small objects. 
A general theoretical framework for the description of Casimir-type forces 
was developed by Lifshitz, Dzyaloshinskii and Pitaevskii~\cite{Lifshitz_SPJ_2_73_1956, Dzyaloshinskii_JETP_10_161_1960, Dzyaloshinskii_SovPhysUsp_4_153_1961}. The framework relies on the theory of fluctuation-induced interactions and can be applied, at least formally, to objects with arbitrary geometries and comprised of a wide range of materials. For a long time since its original development, however, the intrinsic mathematical complexity of the approach meant that accurate calculations were limited to simple geometries, such as planar surfaces, for which analytical evaluations are possible. Due to the increasing relevance for fundamental investigations and technological applications, however, Casimir physics has attracted and continues to attract interest~\cite{Lamoreaux07,CasimirPhysics11}, and this has motivated the development of advanced theoretical and numerical methods for high-accuracy calculations~\cite{CasimirPhysics11, Rodriguez_NPhot_5_211_2011}. 


\begin{figure}[b]
\vspace{-2.5mm}
\begin{overpic}[width=\columnwidth]{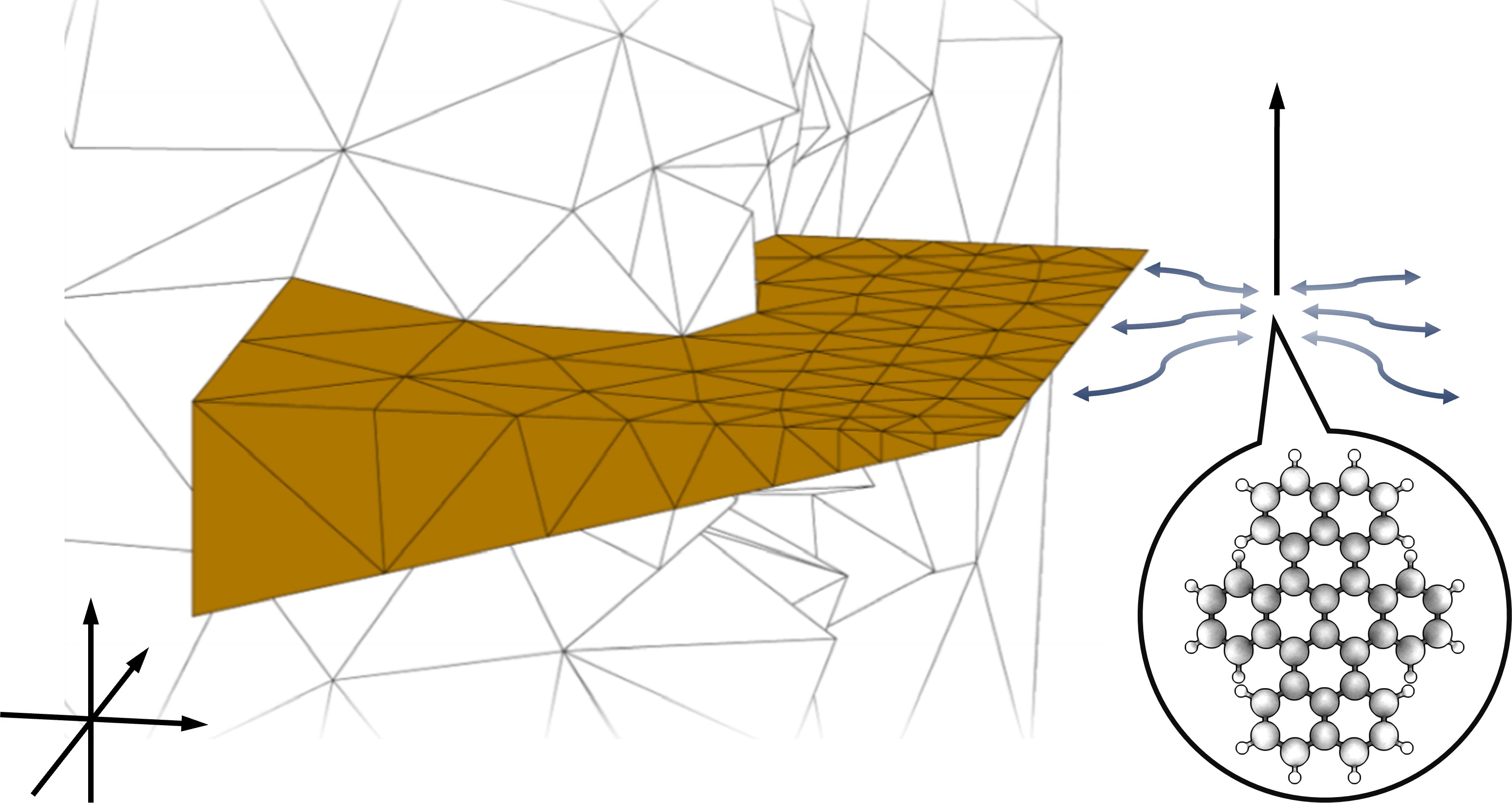}
\put(11,2){$x$}
\put(10.5,7.5){$y$}
\put(7.5,11.5){$z$}
\put(86,44){$F_z$}
\end{overpic}
\\\caption{\label{Fig:prism_sketch} Example material system showing parts of the tetrahedral mesh used for numerical calculations of the Casimir Polder force on a polarizable molecule next to a sharp metallic wedge. Quantum fluctuations of the electromagnetic field, as indicated by blue arrows, lead to a non-trivial repulsive $z$-component of the force in this geometry.}\end{figure}
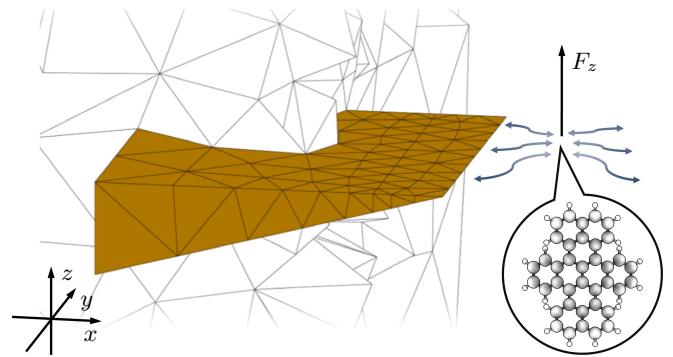

Fluctuation-induced interactions, and Casimir-type force calculations in particular, are historically treated in the frequency domain, where quantum electrodynamic perturbation theory or modal expansions~\cite{Intravaia22a} are possible for  
simple geometries. This was the approach followed by Casimir and Polder to describe the interaction between an atom and a perfectly reflecting surface~\cite{Casimir_PhysRev_73_360_1948} and by Casimir to derive an analytical expression for the force between two infinite and perfectly reflecting planes~\cite{Casimir_Proc_KNAW_51_793_1948}. Similarly, the so-called scattering approach~\cite{Lambrecht_2006,PhysRevD.80.085021,Ingold15a} evaluates Casimir-type forces by a broadband frequency integration over the scattering amplitudes of all the objects in the system. The scattering approach adds flexibility and generality to the evaluation of the interactions. Only recently, however, have semi-analytical calculations using this technique been optimized for specific geometries to the point where it can handle the extreme aspect ratios occurring in typical experiments~\cite{Buscher04, Davids10,Messina15, Hartmann_PRL_119_043901_2017, Antezza20, Schoger22, Emig23}. For general structures, a fully numerical boundary element method implementation of the scattering formulation was presented in Refs. \cite{Reid_PRL_103_040401_2009, Reid_PRA_84_010503_2011, Reid13}. In computational electromagnetism, time-domain calculations offer interesting alternatives to frequency domain formulations and typically show different behaviors in terms of convergence, stability, and memory requirements.  A time-domain method for calculating Casimir forces based on the popular Finite-Difference Time-Domain (FDTD) method was developed and presented in Refs.~\cite{Rodriguez_PRA_76_032106_2007, Rodriguez_PRA_80_012115_2009, McCauley_PRA_81_012119_2009}, and its extension to finite temperatures was discussed in Ref.~\cite{Pan_PRA_83_040503_2011}. 

In this article, we revisit the time-domain approach in order to perform high-accuracy Casimir-Polder force calculations. In comparison to Casimir force calculations, this necessitates the development of a technique to properly treat the dispersive polarizability of the particle. Furthermore, we introduce several additional modifications with respect to previous time-domain approaches. In particular, we do not make a Wick rotation by introduction of an imaginary component to the time-axis, but carry out all calculations in real time by post-processing of the numerical signal. For the numerical calculations, we use the Discontinuous Galerkin Time-Domain (DGTD) method, which offers a number of interesting possibilities compared to FDTD, such as the use of advanced time-stepping schemes and unstructured calculation meshes~\cite{Busch_LaserPhotRev_5_773_2011}. Moreover, DGTD provides an exponential convergence with respect to the order of the polynomial basis functions, 
which is of particular interest for high-accuracy calculations. 
Apart from the relatively easy implementation, the strength of the suggested time-domain approach lies in the immediate availability of a vast catalog of material models developed for simulations of nanophotonic structures. These include high-accuracy fits to experimental material response data~\cite{Wolff_OE_21_12022_2013}, the critical point model to calculate the electromagnetic response of graphene~\cite{Pfeifer22,Werra_Proc_SPIE_9502_2015}, and the hydrodynamic Drude model for investigating effects of nonlocal material responses~\cite{Heinrichs_PRB_7_3487_1973, Ciraci_CPC_14_1109_2013}. In particular, we note that there is a growing interest in the literature for nonlocal material responses in connection with various aspects of Casimir physics~\cite{Svetovoy08a,Mostepanenko21,Klimchitskaya21a,Reiche20, Klimchitskaya21b,Klimchitskaya22}. Previous work on the influence of nonlocal material response in Casimir-type force calculations has been limited to relatively simple geometries~\cite{Esquivel_PRA_68_052103_2003,  Svetovoy06,Bezerra07,Fialkovsky11,Henkel18,Fialkovsky18}, however. In general, therefore, we believe that the approach presented here will be particularly useful for high-accuracy modeling of experiments or nanoscale devices based on Casimir physics with advanced material models.

As a specific application, we consider a configuration where a repulsive Casimir-Polder interaction can occur. The possibilities for repulsive Casimir forces are particularly intriguing, and it was shown already in Refs.~\cite{Dzyaloshinskii_JETP_10_161_1960, Dzyaloshinskii_SovPhysUsp_4_153_1961} by Lifshitz, Dzyaloshinskii, and Pitaevskii, that there can be a repulsive force between objects or particles separated by a liquid, for example. This effect was experimentally demonstrated in Refs~\cite{Meurk97, Feiler08, Munday_Nature_457_45_170_2009, Zhao19}.  Repulsion between objects in vacuum is harder to achieve, and it was shown in Ref.~\cite{Rahi_PRL_070404_2010}, that for purely dielectric or metallic objects in vacuum, there can be no stable equilibria based on Casimir-type forces alone. Nevertheless, repulsion along one specific direction is possible and has been predicted for certain material configurations involving objects with highly anisotropic optical responses, which include a needle near a thin plate with a hole~\cite{Levin_PRL_105_090403_2010} and a wedge~\cite{Milton_JPhysA_45_374006_2012}. The latter is illustrated in Fig.~\ref{Fig:prism_sketch}, which shows an example computational mesh for a wedge and a polarizable molecule oriented along the $z$-direction. If, by other means, the molecule does not rotate and is restricted to move only in the $z$-direction, this geometry leads to a repulsive Casimir-Polder force for positions close to the wedge. To the best of our knowledge, all previous theoretical investigations addressing configurations involving sharp edges have considered either perfect conductors or local material models to describe the optical response of the materials involved~\cite{Buscher04,Davids10,Messina15,Reid_PRL_103_040401_2009, Levin_PRL_105_090403_2010,Maghrebi11, Milton_JPhysA_45_374006_2012,Antezza20,Emig23}. Local response-models, however, are known for giving an inaccurate description of the field distributions in close proximity of an interface and to break down for geometries with sharp edges, where they lead to unphysical divergences in typical scattering calculations~\cite{Andersen_IEEE_AP_26_598_1978}. Nonlocal response-models introduce additional length scales, which effectively round off the sharp corners, and for systems with sufficiently small features, the effect can induce substantial changes in the overall optical response. This raises the question to which extent the repulsive forces are influenced by nonlocal material models and this, in turn, calls for specialized high-accuracy calculation methods, such as the one we present in this work.


The article is organized as follows. In Sec.~\ref{Sec:Formulation} we briefly review the computational framework of Casimir-Polder force calculations along with details pertaining to our particular numerical implementation by use of scattered fields in the DGTD method. In Sec.~\ref{Sec:Example_calculations}, we provide illustrative examples along with detailed discussions of the convergence properties. 
In Sec.~\ref{Sec:Applications} we apply the calculation method to the analysis of the Casimir-Polder force in the vicinity of a sharp wedge described by a nonlocal material response. Finally, we summarize the conclusions of the work in Sec.~\ref{Sec:Conclusion}.

\section{Formulation}
\label{Sec:Formulation}
We consider an electrically polarizable particle near a generic arrangement of bodies made from linear and passive materials, all embedded in vacuum. 
The whole system is in thermal equilibrium. Fluctuations in the particle's dipole moment affect the surrounding electric field just as fluctuations in the surrounding field affect the dipole moment of the particle. Due to the fundamental quantum nature of both particle and field, these fluctuations persist even at vanishing temperature, and even if the fluctuations lead to no average dipole moment or electric field, there is a non-vanishing average interaction energy of the coupled system. Although corrections due to finite temperature can be implemented in the numerical scheme presented in this work, we focus on the case of zero temperature, where the Casimir-Polder force is a pure quantum effect. As a result of electromagnetic scattering from the surrounding bodies, the interaction energy -- which we refer to as the Casimir-Polder energy -- depends on the position of the particle. The interaction may be phrased in terms of second-order correlation functions of the dipole moment and the electric field. In thermal equilibrium the expectation values of these correlation functions are, in turn, related through the fluctuation dissipation theorem \cite{Kubo66,Intravaia_2011} to the imaginary parts of the polarizability tensor $\underline{\alpha}(\omega)$ and the electric field Green tensor $\underline{G}(\mr,\mr',\omega)$, respectively. Consequently, the Casimir-Polder energy may be calculated through an integral over contributions at all angular frequencies as~\cite{Intravaia_2011}
\begin{align}
\mathcal{E}_{\rm CP}(\mr) 
= 
-\frac{\hbar\mu_0}{2\pi}\mathrm{Im} \int_0^\infty
\omega^2\text{Tr}\big[\underline{\alpha}(\omega)
\underline{G}(\mr,\mr,\omega)\big] \ud\omega,
\label{Eq:casPolder}
\end{align}
in which $\hbar$ is the reduced Planck constant, $\mu_0$ is the permeability of free space, and ``$\text{Tr}$'' denotes the trace of the $3\times 3$ matrices describing the tensors and their product.

The electric field Green tensor represents the full retarded electromagnetic response at the position $\mr$ due to an oscillating point dipole source at $\mr'$ and, therefore, contains all information about the scattering properties of the system. For local and isotropic materials described by relative permittivities and permeabilities $\epsilon_\text{R}(\mr,\omega)$ and $\mu_\text{R}(\mr,\omega)$, respectively, it solves the equation 
\begin{multline}
\label{GreenTensors}
\left[\nabla \times \frac{1}{\mu_\text{R}(\mr,\omega)}\nabla \times
\;- \frac{\omega^2}{\text{c}^2} \epsilon_\text{R}(\mr, \omega)\right] \underline{G}(\mr,\mr',\omega) \\
=
\delta(\mr - \mr') \underline{\rm I},
\end{multline}
where $\underline{\rm I}$ denotes the identity tensor and $\text{c}$ is the speed of light in vacuum. In addition to Eq.~\eqref{GreenTensors}, each column of $\underline{G}(\mr,\mr',\omega)$ obeys the same boundary conditions as the electric field at material interfaces as well as a suitable radiation condition to ensure light propagation away from the source region at large distances~\cite{Kristensen20}.
We note that Green tensors can be defined also for nonlocal media, despite the fact that Eq.~\eqref{GreenTensors} then takes a more convoluted form involving integro-differential operators. In general, $\underline{G}(\mr,\mr',\omega)$ can be split into a background contribution, which represents light emission in a homogeneous background medium, and a scattered part $\underline{G}^\text{S}(\mr,\mr',\omega)$, which accounts for all reflections due to the spatial distribution of material in the surroundings. Since for $\mr=\mr'$ the background contribution is independent of position, and since the Casimir-Polder force is given as the negative gradient of the Casimir-Polder energy, 
\begin{align}
\mathbf{F}_\text{CP} (\mr)  = -\nabla\mathcal{E}_{\rm CP}(\mr),
\label{Eq:Force_CP}
\end{align}
the constant background leads to no net contribution to the force. As a consequence, in practical calculations we can replace the Green tensor in Eq.~\eqref{Eq:casPolder} with $\underline{G}^\text{S}(\mr,\mr,\omega)$. 

\subsection{Numerical implementation}
\label{Sec:Numerical_implementation}

Throughout the calculations, we consider a frame in which the polarizability tensor is diagonal. For the method described in the main text this condition is not restrictive, but it can simplify the evaluation and the computational cost. In general, therefore, we set
\begin{align}
[\underline{\alpha}(\omega)]_{ij}=\alpha_{i}(\omega)\delta_{ij},
\end{align}
where $i,j\in\{x,y,z\}$, and the detailed expression for $\alpha_i(\omega)$ depends on the particular model used to describe the behavior of the particle. Given its microscopic nature, the internal dynamics of the particle can often be deduced analytically. The scattering properties of the environment, on the other hand, are described by the Green tensor, which is known analytically only for simple geometries such as spheres or half planes.
From a computational point of view, therefore, the most resource-intensive task in evaluating the Casimir-Polder energy for arbitrary geometries is the numerical calculation of $\underline{G}^\text{S}(\mr,\mr,\omega)$, which amounts to solving the Maxwell equations for the electric field with a point source. To this end, we consider a point source current density of the form $\mJ(\mr,\omega) = \delta(\mr-\mr')J(\omega)\me_j$, where $J(\omega)$ denotes the frequency dependence and $\me_{j}$ is a unit vector in the $j$-direction. Comparing Eq.~(\ref{GreenTensors}) to the wave equation for the electric field with this particular source, 
\begin{align}
\left[\nabla \times \frac{1}{\mu_\text{R}(\mr,\omega)}\nabla \times
\;- \frac{\omega^2}{\text{c}^2} \epsilon_\text{R}(\mr, \omega)\right] \mE(\mr,\omega)
=
\text{i}\omega\mu_0\mJ(\mr,\omega),
\label{Eq:electric_field_eq_freq_dom}
\end{align}
it follows that the $ij$'th component of the Green tensor can be readily obtained from the $i$'th component of the electric field as
\begin{align}
[\underline{G}(\mr,\mr',\omega)]_{ij}= -\text{i}\frac{E_{i}(\mr,\omega;\mr',\me_j)}{\omega\mu_0 J(\omega)},
\label{Eq:G_from_E}
\end{align}
where the dependence of the electric field on the source location $\mr'$ and source dipole orientation $\me_j$ is included explicitly. In the practical calculations below, we will not include them, since we always have $\mr=\mr'$ and $i=j$. 

Comparing Eqs.~\eqref{Eq:casPolder} and \eqref{Eq:G_from_E}, we are particularly interested in the field at the source position. The electric field is known to diverge at point sources, and this leads to a divergence of the Green tensor in the limit $\mr\rightarrow\mr'$. The divergence enters only in the background Green  tensor, and as a consequence it does not affect the Casimir-Polder force. Nevertheless, the attempt to resolve divergent fields in numerical calculations with a finite number of non-divergent basis functions can negatively impact the accuracy. In practical calculations, therefore, we use a scattered-field formulation of Maxwell's equations, as discussed in Appendix~\ref{App:ScatteredFieldFormulation}. This effectively removes the divergence of the field altogether and enables a direct calculation of $\underline{G}^\text{S}(\mr,\mr,\omega)$ by use of Eq.~\eqref{Eq:G_from_E}. With these elements in place, we follow the elegant approach of Refs.~\cite{Rodriguez_PRA_80_012115_2009, McCauley_PRA_81_012119_2009} and rewrite Eq.~(\ref{Eq:casPolder})  
in the form of a convolution as 
\begin{align}
\mathcal{E}_{\rm CP}(\mr)&=- \sum_{i} \hbar \int_0^\infty  \mathrm{Im}[g_{i}(-t)]E^\text{S}_{i}(\mr,t)\,\ud t
\label{Eq:F_j_TimeDom}.
\end{align}
in which $E^\text{S}_{i}(\mr,t)$ denotes the $i$'th component of the scattered electric field generated from a point source current density located at $\mr'=\mr$ and orientated along the direction $\me_j=\me_i$, which also defines the diagonal frame for the polarizability. 
The function $g_{i}(t)$ is the Fourier transform of the function
\begin{align}
g_{i}(\omega)=-\text{i}\alpha_i(\omega)\frac{\omega}{J(\omega)}\Theta(\omega), 
\label{Eq:g_of_omega}
\end{align} 
in which $\Theta(\omega)$ denotes the Heaviside step function.

The freedom in temporal shape of the source field provides interesting options for designing a suitable input field.
In this work, we use a source of the form
\begin{align}
J(t) = J_0\big[4(\gamma t)^3-(\gamma t)^4\big]\text{e}^{-\gamma t}\Theta(t),
\label{Eq:Source_type_1}
\end{align}
which has zero integral and is sufficiently well behaved at small times to be compatible with the scattered field formulation, cf. Appendix~~\ref{App:ScatteredFieldFormulation}. 
The decay of the source acts to reduce the temporal extent of the effective driving field, which substantially lowers the magnitude of the scattered field at times larger than a certain cutoff time set by $\gamma$.
For this particular choice of $J(t)$, the factor $\omega/J(\omega)$ in Eq.~\eqref{Eq:g_of_omega} takes on a polynomial form, which simplifies the transformation of $g_{i}(\omega)$ to the time-domain; in some cases of interest the calculation can be handled analytically, as discussed in Appendix~\ref{App:Calculation_of_g}.
In contrast to Refs.~\cite{Rodriguez_PRA_76_032106_2007, Rodriguez_PRA_80_012115_2009, McCauley_PRA_81_012119_2009}, we do not perform a Wick rotation to artificially damp the oscillations in the electromagnetic response of the source. Rather, we use the physical values of $E_i^\text{S}(\mr,t)$ and rely on the damping induced by the source and the kernel function $\mathrm{Im}[g_{i}(-t)]$to perform the integral in Eq.~\eqref{Eq:F_j_TimeDom} for different particle positions. In a final step, we can calculate the force on the particle by numerical evaluation of the gradient in Eq.~\eqref{Eq:Force_CP}. Since the formulation is fundamentally in terms of energy, however, for most of the examples in this article we shall focus on the numerical evaluation of Eq.~(\ref{Eq:F_j_TimeDom}).
As an additional remark to the formulation, we note that the direct incorporation of material dispersion into time-domain algorithms is rather inefficient, since 
frequency-dependent permittivities or permeabilities lead to numerically demanding integro-differential equations for the temporal dynamics. As discussed in Appendix~\ref{App:ScatteredFieldFormulation}, this problem is circumvented by encoding the material dynamics in separate time-dependent auxiliary differential equations which are solved together with the Maxwell equations~\cite{Taflove2005}.

For numerical calculations we use the DGTD method, which is a finite-element based method specifically designed to solve partial differential equations in conservation form. In terms of the spatial discretization, it has the advantage of being able to use unstructured meshes partitioning space into several, not necessarily equally sized elements. 
This is especially beneficial when modeling irregularly shaped objects, which can only be poorly approximated by discretization on a regular grid. In each of the elements, the electric and magnetic field components are expanded in a local polynomial basis of Lagrange polynomials, in which the expansion coefficients immediately correspond to field values on local grid points within one element. In contrast to classical finite-element methods, there is no overlap between basis functions of adjacent elements. Since this means that all elements can be treated individually, the computation is easily parallelizable and has relatively low memory requirements. The coupling between adjacent elements is then recovered via the introduction of a so-called numerical flux~\cite{Hesthaven08, Busch_LaserPhotRev_5_773_2011}. In addition to refining the mesh, adjusting the polynomial order $p$ of the basis functions can be used to control the error connected with spatial discretization. Letting $h$ denote the average element size, the discretization error can be shown to behave as $\mathcal{O}(h^{p+1})$~\cite{Busch_LaserPhotRev_5_773_2011} for sufficiently small $h$ or large $p$, and assuming piecewise planar interfaces. For curved interfaces, the error will eventually be limited by the geometrical approximation of the mesh, leading to an error that scales as $\mathcal{O}(h^{2})$ for spheres represented by straight-sided elements, for example. In such cases, $p$-adaptivity can be restored by the use of curvilinear elements~\cite{Busch_LaserPhotRev_5_773_2011,Viquerat15}. All together, the resulting spatial discretization gives rise to a set of first order ordinary differential equations in time, which can in principle be solved by one of several advanced time stepping algorithms known in the literature. Here, we use a 4th-order Low-Storage Runge-Kutta method with 14 stages~\cite{Niegemann12}. 
The time-domain calculation then provides the electric field at all discrete times and at specified positions. 

For practical calculations, we must choose a length scale, and throughout this work we use $L_0=100\,$nm as a reference.  
From Eq.~(\ref{Eq:casPolder}) we can then define
\begin{align}
\mathcal{E}_0 = \frac{\hbar\mu_0\text{c}^3}{2\pi}\frac{\alpha_0}{L_0^4}
\label{Eq:U_0}
\end{align}
where 
$\alpha_0$ is a constant and scalar polarizability, as a convenient unit of energy. 

\section{Example calculations}
\label{Sec:Example_calculations}
In this section, we illustrate the methodology in setting up and carrying out the calculations in practice by considering the general problem of 
an isotropic polarizable particle above an infinite surface. This simple configuration enables a semi-analytical description of the interaction which will be used as reference to analyze the convergence behavior. For simplicity, we neglect the magnetic properties of all involved objects. In Sec.~\ref{Sec:Casimir_Polder_force}, we consider the case of a simple gold surface and provide a detailed account of the numerical calculation scheme as well as a thorough analysis of the convergence. Next, in Sec.~\ref{Sec:Non_local_material_response}, we demonstrate   
the compatibility of the method with advanced material models by extending the calculations to the case of a nonlocal material model, thereby setting the stage for the application example in Sec.~\ref{Sec:Applications}.  

\subsection{Particle above a gold surface}
\label{Sec:Casimir_Polder_force}
We consider the Casimir-Polder energy of a polarizable particle in vacuum at a distance $z$ above an infinitely extended gold surface described by the Drude model
\begin{align}
\epsilon_\text{D}(\omega) = 1 - \frac{\omega_\text{pl}^2}{\omega^2 + \text{i}\Gamma\omega},
\label{Eq:Epsilon_Drude}
\end{align}
where $\omega_\text{pl}$ is the plasma frequency and $\Gamma$ the dissipation rate. Throughout this article, we use the values for gold reported in Ref.~\cite{Intravaia_Nature_Comm_4_2515_2013} by setting $\omega_\text{pl}=8.39\,\text{eV}$ and $\Gamma=0.0434\,\text{eV}$, corresponding to $\omega_\text{pl} = 1.2747\times10^{16}\text{s}^{-1}$ and $\Gamma = 6.5936\times10^{13}\text{s}^{-1}$. 
For the particle, we consider an isotropic dispersive polarizability of the form $\underline{\alpha}(\omega)=\alpha(\omega)\underline{\rm I}$, where 
\begin{equation}
\alpha(\omega) = \alpha_0\frac{\omega_\text{a}^2}{\omega_\text{a}^2 - \omega^2 - \text{i} \gamma_\text{a} \omega},
   \label{Eq:alpha_ho}
\end{equation}
in which $\alpha_0$ is the static polarizability, $\omega_\text{a}$ is the resonance frequency, and $ \gamma_\text{a}$ is the damping rate. 
For the calculations in this section, we use parameters corresponding to a Rubidium-87 atom by choosing $\omega_\text{a}=1.6\,\text{eV}$ as the resonance frequency and $ \gamma_\text{a} =2.5 \cdot 10^{-8}\,\text{eV}$ as the linewidth of the relevant atomic transition \cite{Steck2015}.

As detailed in Appendix~\ref{App:Calculation_of_g}, the mathematical form of the polarizability enables an analytical evaluation of the corresponding kernel function $g_i(t)=g(t)$, which in this case is independent of direction because of the isotropic polarizability. From the analytical result, it follows that the time-dependence of the source in combination with the expression for the polarizability gives rise to a temporal behavior for $\mathrm{Im}[g(-t)]$ which diverges as $\mathcal{O}(t^{-4})$ at short times and goes to zero as $\mathcal{O}(t^{-1})$ at long times.
\begin{figure}[htb]
\centering
\begin{overpic}[width=0.9\columnwidth]
{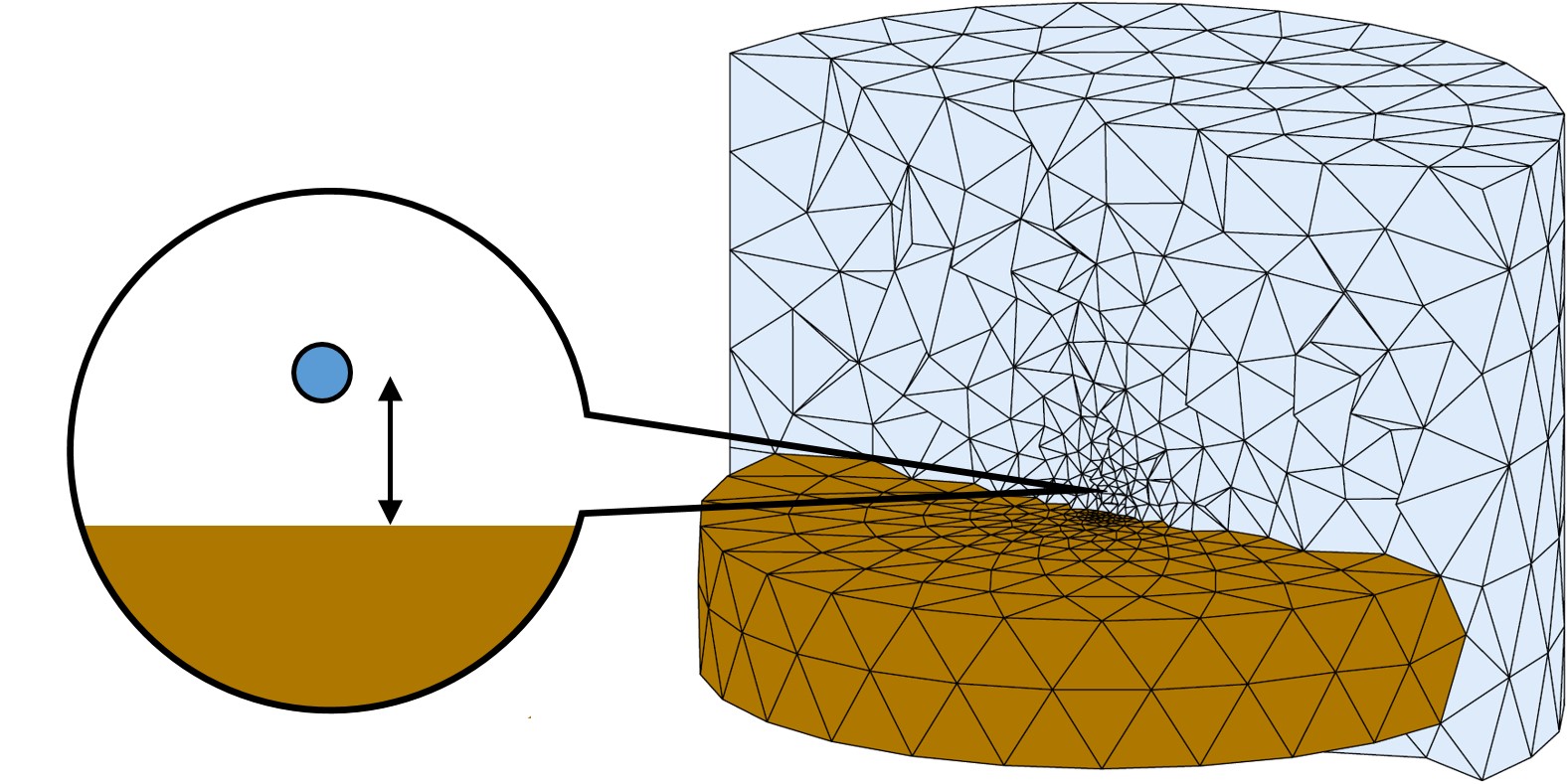}
\put(27,20){$z$}
\end{overpic}
\caption{\label{Fig:cylEmbedded_Hcyl_1p5_Hair_4p0_Pad_0p5_R_4p0_d1_0p2_h_0p1_v2}Left: Model setup of a polarizable particle at a distance $z$ above a gold surface. 
Right: Example calculation mesh consisting of a circular disk of radius $R=400\,$nm embedded in vacuum.  
The mesh is finer at positions close to the source position.} 
\end{figure}

To enable a numerical solution by the DGTD method, we first set up a tetrahedral calculation mesh of the geometry. For the example problem at hand, the mesh consists of a circular disk of radius $R$ much larger than the atom-surface separation embedded in a background of vacuum, as shown in Fig.~\ref{Fig:cylEmbedded_Hcyl_1p5_Hair_4p0_Pad_0p5_R_4p0_d1_0p2_h_0p1_v2}. The thickness of the disk is $H_\text{disk}=150\,$nm. The calculation domain extends above the disk to a height of $H=R$, and there is an additional padding layer of vacuum of thickness $H_\text{pad}=50\,$nm surrounding the entire domain. In this way, the radius of the disk defines the calculation domain size. 
Apart from a fixed circular and flat mesh surface at the height of $z_0=20\,$nm above the central part of the disk to help interpolation, the mesh is unstructured and consists of relatively small elements of side length $h\approx1$\,nm in the central region, and elements with side length as large as $h\approx100$\,nm near the calculation domain boundary.  
Whereas the time-dependent Maxwell equations are solved in the entire domain, the auxiliary differential equation describing the current density in the Drude model, as discussed in Appendix~\ref{App:DrudeModel}, is defined and solved only inside the disk. In this particular example, 
the translational symmetry of the system means that the Green tensor depends only on the distance to the surface. Moreover, the rotational symmetry in combination with the isotropic polarizability means that only two unique contributions to the trace remain. We can therefore write the Casimir-Polder energy as
\begin{align}
\mathcal{E}_{\rm CP}(z) = 2\mathcal{E}_{\|}(z) + \mathcal{E}_{\perp}(z),
\label{Eq:E_CP_par_perp_contribs}
\end{align}
in which the two contributions, denoted by ``$\|$'' and ``$\perp$'' and corresponding to orientations parallel and perpendicular to the surface, respectively, must be calculated individually. 

\begin{figure}[htb]
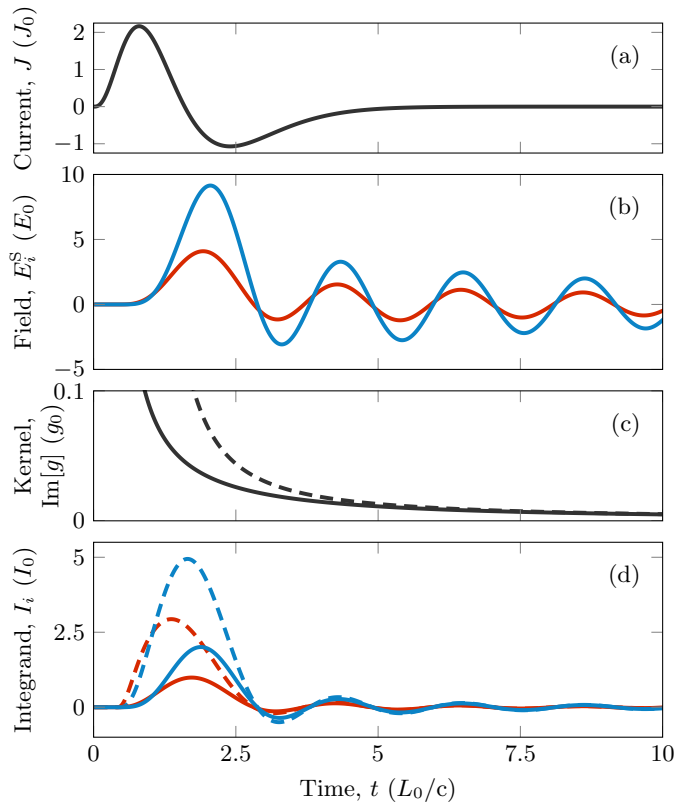

\flushright
\input{fig3a.tex}
\\[2.5mm]
\input{fig3b.tex}
\\[2.5mm]
\input{fig3c.tex}
\\[2.5mm]
\input{fig3d.tex}
\\[7mm]
\caption{\label{Fig:CasimirPolder_full_J}Details of the elements entering the numerical calculation of the Casimir-Polder energy. {(a)}: Temporal current profile corresponding to Eq.~\eqref{Eq:Source_type_1} with $\gamma = 2.5 \text{c}/L_0$. {(b)}: Resulting scattered electric field components $E^\text{S}_i(z_0,t)$ in units of $E_0=J_0/\epsilon_0\text{c}L_0^2$ and evaluated at the source position. For times $t<2z_0/\text{c}$, the signals are identically zero because of the finite time required to travel from the source to the surface and back. 
{(c)}: The kernel function $\text{Im}[g(-t)]$ in units of $g_{0}=\alpha_{0}\text{c}^3/J_0L_0^3$. {(d)}: Integrand $I_i(z_0,t)=\mathrm{Im}[g(-t)]E^\text{S}_i(z_0,t)$ from Eq.~\eqref{Eq:F_j_TimeDom} in units of $I_0=\mathcal{E}_0\text{c}/\hbar L_0$. The red and blue curves in panels b and d correspond to $i=x,y$ and $i=z$, respectively. Dashed lines in panels c and d show as reference the results using a constant polarizability of magnitude $\alpha_0$.}
\end{figure}
For the calculations, we choose a source current with $\gamma= 2.5\text{c}/L_0$, as illustrated in panel (a) of Fig.~\ref{Fig:CasimirPolder_full_J}, and using the scattered field formulation, we evolve the scattered electromagnetic fields in time and record $E_i^\text{S}(z,t)$ at the source point position above the center of the surface. The response to the source is generally oscillatory with a frequency which depends on the details of the source, the distance from the surface, and the surrounding materials.
The oscillations will be damped because of loss in the material and radiative loss to the environment, but the damping is relatively slow, as seen in panel (b) of Fig.~\ref{Fig:CasimirPolder_full_J}. 
Next, we multiply by the function $\mathrm{Im}[g(-t)]$, which enhances the signal at short times and damps it at larger times, as illustrated in panel (c).  The integrand in Eq.~\eqref{Eq:F_j_TimeDom} is therefore much more well behaved than the original signal, as seen in panel (d). To highlight the effects of the dispersive polarizability, the results are shown also for the case of a constant polarizability of magnitude $\alpha_0$. 
Importantly, the divergence of $\mathrm{Im}[g(-t)]$ at short times is cured by the fact that the scattered field is zero for times $t<2z/\text{c}$ as a consequence of retardation. Finally, to calculate the Casimir-Polder energy, we use a trapezoidal integration to approximate the integral in Eq.~\eqref{Eq:F_j_TimeDom} from $t=0$ to a given cut-off time $t_{\rm cut}$. 
\begin{figure}[htb]
\flushright
%
%
\definecolor{mycolor1}{rgb}{0.84375,0.16016,0.00000}%
\definecolor{mycolor2}{rgb}{0.04297,0.51562,0.77734}%
\definecolor{mycolor3}{rgb}{0.39062,0.39062,0.39062}%
\begin{tikzpicture}
\draw[draw=none, use as bounding box](0,0) rectangle (0.875\columnwidth,0.3\columnwidth);
\begin{axis}[%
compat=1.3, 
width=.875\columnwidth,
height=0.3\columnwidth,
scale only axis,
axis on top,
xmin=15,
xmax=55,
xtick={20, 30, 40, 50},
ymin=-2.5,
ymax=0,
ytick={-2, -1,  0},
xlabel style={align=center},
xlabel={Distance to interface, $z$ (nm)},
ylabel style={align=center},
ylabel={Energy, $\mathcal{E}_{\|/\perp}$ ($\mathcal{E}_0$)},
axis background/.style={fill=white}
]
\addplot [color=mycolor1, line width=1.5pt, forget plot]
  table[row sep=crcr]{%
10	-9.82120331707642\\
11	-7.36315525891298\\
12	-5.65916560111144\\
13	-4.4411998598927\\
14	-3.54783417044967\\
15	-2.8779002344468\\
16	-2.36580291115122\\
17	-1.96774673570053\\
18	-1.65373402446291\\
19	-1.40275052234027\\
20	-1.199778562757\\
21	-1.0338930286426\\
22	-0.897017854626208\\
23	-0.783096107673924\\
24	-0.687525146830325\\
25	-0.606765312551973\\
26	-0.538064416221842\\
27	-0.479260871151743\\
28	-0.428641093978315\\
29	-0.384834914721608\\
30	-0.346737971421141\\
31	-0.313453505708683\\
32	-0.284248271140809\\
33	-0.258518819978363\\
34	-0.235765500212888\\
35	-0.215572235347879\\
36	-0.197590680144191\\
37	-0.181527715628943\\
38	-0.167135512440442\\
39	-0.154203584304839\\
40	-0.142552394477541\\
41	-0.132028182091428\\
42	-0.12249875283467\\
43	-0.113850036492452\\
44	-0.105983257792598\\
45	-0.0988126003999357\\
46	-0.092263269488887\\
47	-0.0862698780432454\\
48	-0.0807750973227275\\
49	-0.0757285238608017\\
50	-0.0710857247095396\\
51	-0.0668074300189302\\
52	-0.0628588478785909\\
53	-0.0592090809995317\\
54	-0.0558306285325028\\
55	-0.0526989593070876\\
56	-0.049792145186118\\
57	-0.0470905451827175\\
58	-0.0445765325754682\\
59	-0.0422342585538364\\
60	-0.0400494469884673\\
};
\addplot [color=mycolor2, line width=1.5pt, forget plot]
  table[row sep=crcr]{%
10	-19.2056385209724\\
11	-14.3490694454448\\
12	-10.9891577852026\\
13	-8.59266404185633\\
14	-6.83873041158346\\
15	-5.52645725891694\\
16	-4.52571565074301\\
17	-3.7497140265151\\
18	-3.13906942579811\\
19	-2.65222961113527\\
20	-2.25953462294672\\
21	-1.93943362607707\\
22	-1.67601586648064\\
23	-1.4573639740805\\
24	-1.27443404399102\\
25	-1.12028036320588\\
26	-0.989509988076867\\
27	-0.877893319275723\\
28	-0.782082261208624\\
29	-0.699403679107321\\
30	-0.627706277982597\\
31	-0.565245863142188\\
32	-0.510598500492962\\
33	-0.462594179149426\\
34	-0.420265693977556\\
35	-0.382808934479724\\
36	-0.349551798408438\\
37	-0.319929681591021\\
38	-0.293466021619218\\
39	-0.269756754395312\\
40	-0.248457821435935\\
41	-0.229275071597722\\
42	-0.211956053933165\\
43	-0.196283313101271\\
44	-0.18206888537775\\
45	-0.169149759171882\\
46	-0.157384114369258\\
47	-0.146648193649953\\
48	-0.136833689021473\\
49	-0.127845550256896\\
50	-0.119600140306077\\
51	-0.112023677225124\\
52	-0.105050913631691\\
53	-0.0986240138125799\\
54	-0.0926915958984774\\
55	-0.0872079123715973\\
56	-0.0821321468892059\\
57	-0.0774278092246644\\
58	-0.0730622132311444\\
59	-0.0690060252650545\\
60	-0.0652328725793117\\
};
\addplot [color=mycolor3, line width=1.5pt, only marks, mark=o, mark options={solid, mycolor3}, forget plot]
  table[row sep=crcr]{%
20	-1.22200496487783\\
30	-0.359557295933457\\
40	-0.148214685633531\\
50	-0.0724931114186896\\
};
\addplot [color=mycolor3, line width=1.5pt, only marks, mark=o, mark options={solid, mycolor3}, forget plot]
  table[row sep=crcr]{%
20	-2.48001984702221\\
30	-0.628607486302388\\
40	-0.239625812711363\\
50	-0.120992621281393\\
};
\addplot [color=black!50!mycolor3, line width=1.5pt, only marks, mark=square, mark options={solid, black!50!mycolor3}, forget plot]
  table[row sep=crcr]{%
20	-1.19595929734316\\
30	-0.346547643425173\\
40	-0.142538695323528\\
50	-0.0710015132455696\\
};
\addplot [color=black!50!mycolor3, line width=1.5pt, only marks, mark=square, mark options={solid, black!50!mycolor3}, forget plot]
  table[row sep=crcr]{%
20	-2.26570849415167\\
30	-0.629448689473978\\
40	-0.248602342233865\\
50	-0.119387363394442\\
};
\end{axis}
\end{tikzpicture}%
\\[7mm]
\caption{\label{Fig:CasimirPolder_full_FIGold_100nm_varPos}Contributions to the Casimir-Polder energy as a function of distance above a gold surface. 
The red and blue curves show $\mathcal{E}_\|(z)$ and $\mathcal{E}_\perp(z)$, respectively, and light circular and dark square markers show the numerical results obtained when using the same computational mesh, but first and second order polynomial basis functions. Due to the exponential convergence, the differences between results using second and third order basis functions are not visible on this scale. 
}
\end{figure}
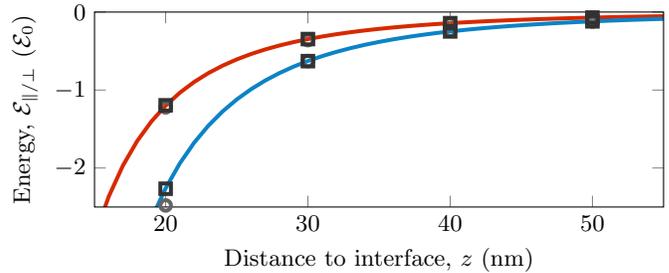
Whereas the temporal current profile and the resulting calculated fields are independent of the particle's properties, the specific functional form of the polarizability affects the function $\mathrm{Im}[g_{i}(-t)]$ and thereby the resulting integrand.  
This effectively means that the details of the particle are accounted for only in post-processing, such that a given numerical simulation of the electromagnetic fields -- which generally represents the resource-intensive part of the solution process -- can be subsequently used to study the Casimir-Polder energy for different particles or atom species at practically no additional computational cost. 
Figure~\ref{Fig:CasimirPolder_full_FIGold_100nm_varPos} shows the numerical results for the Casimir-Polder energy as a function of distance from the surface along with the result of a reference calculation obtained directly from Eq.~\eqref{Eq:casPolder} by integrating the known expression of the Green tensor in the frequency domain~\cite{Wylie84,Tomas95,Paulus_PRE_62_2000}. 
The numerical calculations reproduce the reference values, and the relative errors decrease with higher polynomial order of the basis functions as expected. 
In Fig.~\ref{Fig:CasimirPolder_full_FIGold_100nm_varPos}, this is especially noticeable at short distances where the magnitude of the energy is largest.

\subsubsection{Convergence}
\label{Sec:Convergence_CasPolder}

In addition to errors stemming from the time integration, the quality of the numerical result is limited by the accuracy of the space and time discretization in the DGTD method as well as residual reflections from the numerical calculation domain boundary. Below, we describe each of these sources of error individually. 

\begin{figure}
\flushright
\input{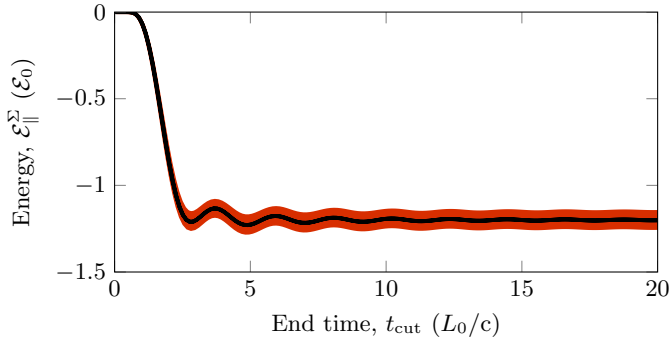}
\\[7mm]
\caption{\label{Fig:CasPolder_FIGold_100nm_z_20nm_Ftss_vs_time_orders_1_5}
Trapezoidal cumulative sum $\mathcal{E}^\Sigma_{\|}(z_0,t_{\rm cut})$ as a function of end time $t_{\rm cut}$ and in units of $\mathcal{E}_0$. The cumulative sum is approximating the $i=x,y$ components of the integral in Eq.~\eqref{Eq:F_j_TimeDom}. The widths of the colored areas represent the extrema of the curves as calculated for 100 nominally identical positions with basis function polynomial order $p=1$ (red) and $p=5$ (black).
} \end{figure}

\emph{Time integration}: As a practical way of estimating the true value of the integral in Eq.~\eqref{Eq:F_j_TimeDom}, we calculate the trapezoidal cumulative sum $\mathcal{E}^\Sigma(z,t_{\rm cut})$ corresponding to the trapezoidal integration of Eq.~\eqref{Eq:F_j_TimeDom} up to $t=t_{\rm cut}$. 
The cumulative sum inherits the oscillatory behavior of the original field, but 
the oscillations occur around the numerical estimate of the true value, when disregarding other sources of error. For a given end time, we obtain a numerical estimate of the true value by calculating the zeroth-order term in the Fourier expansion of $\mathcal{E}^\Sigma(z,t_{\rm cut})$ evaluated from the last few full periods of oscillations. 
The difference between the extrema and the estimated true value can serve as a very conservative estimate of the numerical error due to the time integration. This estimate becomes better for larger values of $t_{\rm cut}$ at the expense of longer calculation times. In practice, however, the error can be several orders of magnitude lower than this conservative estimate. In the case of $z=z_0$, Fig.~\ref{Fig:CasPolder_FIGold_100nm_z_20nm_Ftss_vs_time_orders_1_5} shows the cumulative sum as a function of the end time, for which the oscillations around the reference value are clearly visible.

\emph{Spatial discretization}: As discussed in Sec.~\ref{Sec:Numerical_implementation}, the discontinuous nature of the DGTD method leads to an interesting convergence behavior. The numerical flux enforces the correct behavior of the fields across the boundaries between different mesh elements, but this enforcement is in the weak sense only. Indeed, for large mesh elements or low polynomial order, this may lead to numerical errors in the calculated fields at material interfaces or boundaries between different elements within the same material. To assess the convergence, therefore, it is reasonable to look at the statistical properties of the error in different, nominally identical calculations. The red area in 
Fig.~\ref{Fig:CasPolder_FIGold_100nm_z_20nm_Ftss_vs_time_orders_1_5}  represents the variation in 100 calculations,  
each corresponding to different positions close to the center, but all with the same distance to the surface, and calculated using basis functions of order $p=1$. The effect of increasing the basis function order is to narrow the distance between the curves, as illustrated by the black area, which corresponds to the same positions in the same mesh, but calculated using basis functions of order $p=5$.

\emph{Reflections from the boundary}: Because of the typical fast decay of Casimir-type forces, the influence of the calculation domain boundary on the evaluation of the Casimir-Polder energy in Eq.~\eqref{Eq:F_j_TimeDom} can be effectively controlled by varying the size of the domain. This largely alleviates the need for more sophisticated techniques to deal with unwanted reflections from the boundary, such as perfectly matched layers, since the physics of the problem itself leads to vanishing influence of distant interfaces. This is very different from the calculation of other physical quantities in nanophotonics, which can depend dramatically on residual reflections from the domain boundaries. Still, given the accuracy of the evaluation, we do see an influence of small reflections. Therefore, in practice we use first-order Silver-M{\"u}ller outgoing wave boundary conditions, since they can be incorporated in the numerical scheme with no additional calculation cost~\cite{Busch_LaserPhotRev_5_773_2011}. 
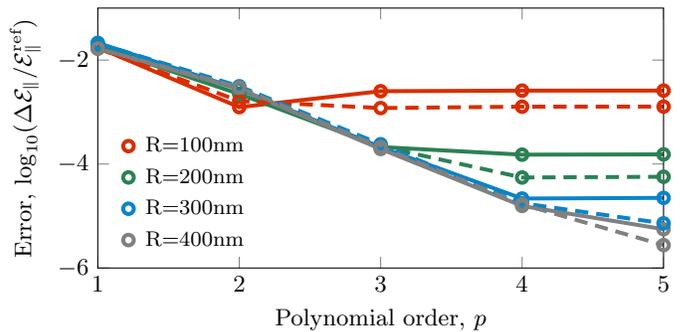
\begin{figure}
\flushright
%
%
\definecolor{mycolor1}{rgb}{0.84706,0.18039,0.00000}%
\definecolor{mycolor2}{rgb}{0.16863,0.50588,0.33725}%
\definecolor{mycolor3}{rgb}{0.04314,0.51765,0.78039}%
\definecolor{mycolor4}{rgb}{0.50196,0.50196,0.50196}%
\begin{tikzpicture}
\draw[draw=none, use as bounding box](0,0) rectangle (0.87\columnwidth,0.4\columnwidth);
\begin{axis}[%
compat=1.3, 
width=.87\columnwidth,
height=0.4\columnwidth,
scale only axis,
axis on top,
xmin=1,
xmax=5,
xtick={1, 2, 3, 4, 5},
ymin=-6,
ymax=-1,
xlabel style={align=center},
xlabel={Polynomial order, $p$},
ylabel style={align=center},
ylabel={Error, $\log_{10}(\Delta\mathcal{E}_{\|}/\mathcal{E}_{\|}^\text{ref})$},
axis background/.style={fill=white},
legend style={at={(0.03,0.03)}, anchor=south west, legend cell align=left, align=left, fill=none, draw=none,font=\footnotesize,row sep=1pt}
]
\addplot+[color=mycolor1, line width=1.5pt, mark=o, mark options={solid, mycolor1}, forget plot, legend image post style={only marks}]
  table[row sep=crcr]{%
1	-1.75799205805384\\
2	-2.90587948479728\\
3	-2.59881126188801\\
4	-2.58772760415293\\
5	-2.58693971046663\\
}; \addlegendentry{\;R=100nm}

\addplot+[color=mycolor2, line width=1.5pt, mark=o, mark options={solid, mycolor2}, forget plot, legend image post style={only marks}]
  table[row sep=crcr]{%
1	-1.73860462956773\\
2	-2.65240774562219\\
3	-3.67264030228479\\
4	-3.81969095655525\\
5	-3.81423189854795\\
}; \addlegendentry{\;R=200nm}

\addplot+[color=mycolor3, line width=1.5pt, mark=o, mark options={solid, mycolor3}, forget plot, legend image post style={only marks}]
  table[row sep=crcr]{%
1	-1.66898072331749\\
2	-2.5584966228262\\
3	-3.67722570728241\\
4	-4.66382485474479\\
5	-4.65110996817559\\
}; \addlegendentry{\;R=300nm}

\addplot+[color=mycolor4, line width=1.5pt, mark=o, mark options={solid, mycolor4}, forget plot, legend image post style={only marks}]
  table[row sep=crcr]{%
1	-1.7770881232368\\
2	-2.55704427602468\\
3	-3.71097277471685\\
4	-4.80624035661697\\
5	-5.25345449358323\\
}; \addlegendentry{\;R=400nm}

\addplot+[color=mycolor1, dash pattern={on 5pt off 3pt}, line width=1.5pt, mark=o, mark options={solid, mycolor1}, legend image post style={only marks}]
  table[row sep=crcr]{%
1	-1.773887018558\\
2	-2.78821581206197\\
3	-2.9229093964405\\
4	-2.89654123236675\\
5	-2.89497232558545\\
};

\addplot+[color=mycolor2, dash pattern={on 5pt off 3pt}, line width=1.5pt, mark=o, mark options={solid, mycolor2}, legend image post style={only marks}]
  table[row sep=crcr]{%
1	-1.75521427309613\\
2	-2.53620908511219\\
3	-3.66091282292206\\
4	-4.25915602518056\\
5	-4.24496167045471\\
};

\addplot+[color=mycolor3, dash pattern={on 5pt off 3pt}, line width=1.5pt, mark=o, mark options={solid, mycolor3}, legend image post style={only marks}]
  table[row sep=crcr]{%
1	-1.67959141134799\\
2	-2.50376516022356\\
3	-3.61933083621462\\
4	-4.75329602469429\\
5	-5.14108757699028\\
};

\addplot+[color=mycolor4, dash pattern={on 5pt off 3pt}, line width=1.5pt, mark=o, mark options={solid, mycolor4}, legend image post style={only marks}]
  table[row sep=crcr]{%
1	-1.77844378461839\\
2	-2.52870823801981\\
3	-3.67622826178085\\
4	-4.77577545347032\\
5	-5.55937466428423\\
};

\end{axis}
\end{tikzpicture}%
\\[7mm]
\caption{\label{Fig:CasPolder_FIGold_100nm_z_20nm_log10error_vs_order}Average relative error of the Casimir-Polder energy calculations as a function of basis function polynomial order $p$ and for different calculation domain sizes $R=H$, where $R$ is the radius of the disk and $H$ is the distance from the surface to the upper boundary of the mesh. The figure shows the parallel component $\Delta\mathcal{E}_{\|}(z_0)=\mathcal{E}_{\|}(z_0)-\mathcal{E}^{\rm ref}_{\|}(z_0)$ corresponding to $i=x,y$, and the dashed and solid lines correspond to constant and dispersive polarizabilities, respectively.}
\end{figure}
For the case of $z=z_0$, Fig.~\ref{Fig:CasPolder_FIGold_100nm_z_20nm_log10error_vs_order} shows the average relative error with respect to the reference calculation for $\mathcal{E}_{\|}(z)$ as a function of polynomial order and for different calculation domain sizes. The average was taken over 100 calculations performed on the same mesh by varying the lateral source position through a matrix of $x,y$-coordinates close to the center. As discussed in Sec.~\ref{Sec:Numerical_implementation}, the error in the DGTD method itself is expected to vary exponentially with polynomial order of the basis functions, and this is visible in Fig.~\ref{Fig:CasPolder_FIGold_100nm_z_20nm_log10error_vs_order} in the curves corresponding to large calculation domain sizes.  
The apparent dip in the $R=100\,$nm curve at $p=2$ does not signify an improved performance with these settings, but is likely due to addition of two independent errors from limited discretization and reflections from the boundary which incidentally add up to produce a final result closer to the reference value than the results with larger calculation domains.

\subsection{Nonlocal material response}
\label{Sec:Non_local_material_response}

The simple description of materials in terms of a local dielectric function is known to break down at short length scales where additional physical phenomena become relevant~\cite{FeibelMan_ProgSurfSci_12_287_1982, Mortensen21a}. In this limit, local material models lead to unphysical results, such as divergences at sharp corners and tips~\cite{Andersen_IEEE_AP_26_598_1978}. For metals, the length scales characterizing nonlocal effects are the electron mean free path $\ell$ and the Thomas-Fermi screening length $\lambda_{\rm TF}$. For gold, $\ell$ and $\lambda_{\rm TF}$ are on the order of a few tens of nanometers and a few Angstrom, respectively. The former gives the average length of the electrons' ballistic motion within the metal and mainly describes diffusive mechanisms in the conductor, while the latter describes the size of the screening effects due to a spatial reordering of the charge density and sets the scale for the convective dynamics~\cite{Kittel96,Mortensen21a}. 
Both can modify the interaction with the electromagnetic radiation and in principle must be taken into account in its description.

In nonlocal materials the bulk permittivity is described in terms of a transverse and a longitudinal dielectric function, $\epsilon_\text{t}(\omega,k)$ and $\epsilon_\text{l}(\omega,k)$ respectively \cite{Ford84,Intravaia15a}. Of the many models available to describe the general nonlocal response of a metal~\cite{FeibelMan_ProgSurfSci_12_287_1982, Mortensen21a}, we consider here the so-called hydrodynamic model of the charge and current density~\cite{Heinrichs_PRB_7_3487_1973, Ciraci_CPC_14_1109_2013} as discussed in Appendix~\ref{App:LinHydro}. Indeed, time-domain calculations in general are well suited to handle hydrodynamic models, and the conservation form description underlying the DGTD calculation framework makes it particularly well suited for this task~\cite{Hille_JPCC_120_1163_2016,Wegner23}. We focus on the linearized version of the hydrodynamic model, in which the transverse dielectric function is given by the Drude model in Eq.~\eqref{Eq:Epsilon_Drude}, and only the longitudinal dielectric function is changed. Thus, we set $\epsilon_\text{t}(\omega,k)=\epsilon_\text{D}(\omega)$ and $\epsilon_\text{l}(\omega,k)=\epsilon_{\rm HD}(\omega,k)$, in which 
\begin{align}
\epsilon_{\rm HD}(\omega,k)= 1 - \frac{\omega_\text{pl}^2}{\omega^2+\text{i}\Gamma\omega - \beta^2k^2},
\label{hydroM}
\end{align}
where $\beta$ is the speed of sound in the electron continuum, which is related to the compressibility of the electronic fluid~\cite{Lindhard54}, and $k$ is the magnitude of the electromagnetic field wave vector. In metals, $\beta$ is a function of the Fermi velocity $v_\text{F}$ with the exact expression depending on the particular material model~\cite{Wegner23}. In this work, we take the low-frequency limit $\beta=v_\text{F}/\sqrt{3}$, where the ratio $\beta/\text{c}$ ranges from $1.5\times 10^{-3}$ to $6\times 10^{-3}$ in typical metals. For gold, in particular, we find $\beta/\text{c} \approx 3\times10^{-3}$~\cite{Ashcroft76}. Comparing to Eq.~(\ref{Eq:Epsilon_Drude}), the nonlocal effects are visible as a spatial dispersion governed by $\beta$. 
In its simplicity, this model takes into account the convective process of the electron gas in the metal but we note that it neglects diffusion and other physical phenomena~\cite{FeibelMan_ProgSurfSci_12_287_1982,Mortensen21a}. As in the case of the local Drude model, the auxiliary differential equations describing the hydrodynamic model are defined only within the disk, where the material response is taken to be that of the bulk material governed  by Eq.~(\ref{hydroM}). The model describes the charge density as well as the current density, and the  
introduction of additional dynamical variables are complemented by extra boundary conditions~\cite{Baryakhtar91, Baryakhtar91a}. Specifically, we require the normal component of the current density to vanish at the metal-vacuum interfaces. More details and discussions about the treatment of material interfaces are presented in Appendix~\ref{App:LinHydro}.

To assess the potential of the suggested calculation scheme for use with nonlocal material responses, we consider again the general setup from Sec.~\ref{Sec:Casimir_Polder_force} with a polarizable particle modeled by a dispersive polarizability at a fixed distance of $z=z_0$ above a gold surface, but we now replace the Drude model for the gold with the linearized hydrodynamic model, as detailed in Appendix~\ref{App:LinHydro}. 
Figure~\ref{Fig:CasimirPolderAnalytical_FIGold_100nm_z_20nm_Es_vs_b} shows the numerical results along with independent reference calculations $\mathcal{E}_{\|}^\text{ref}(z_0,\beta)$ based on direct calculation of the Green tensor by taking into account the change in reflection coefficients due to the spatial dispersion of the permittivity~\cite{Mochan_PRB_35_1088_1987,Esquivel_PRA_68_052103_2003}. 
\begin{figure}
\flushright
\input{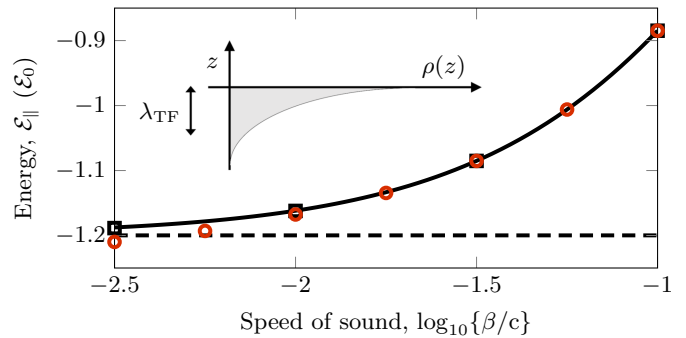}
\\[7mm]
\caption{\label{Fig:CasimirPolderAnalytical_FIGold_100nm_z_20nm_Es_vs_b}Casimir-Polder energy $\mathcal{E}_{\|}(z_0)$ above a gold surface with a nonlocal material model as a function of the speed of sound in the electron continuum $\beta$. The solid line shows the high-precision reference calculation $\mathcal{E}_{\|}^\text{ref}(z_0)$, and red circles show the numerical results obtained using a standard mesh as in Fig.~\ref{Fig:cylEmbedded_Hcyl_1p5_Hair_4p0_Pad_0p5_R_4p0_d1_0p2_h_0p1_v2}. The dashed horizontal line indicates the result using the local Drude model with $\beta=0$. The inset shows schematically the variation of the charge density near the surface with a characteristic thickness of $\lambda_{\rm TF}$. By changing the mesh to introduce an extreme sampling close to the surface, one can resolve the sharp rise in the charge density for small values of $\beta$, as indicated by the black squares.
} 
\end{figure}
Contrary to the case of a local material response, the finite compressibility in the hydrodynamic model describes a smearing out of the charge density below the surface with a characteristic length governed by the Thomas-Fermi screening length. The connection to the local material response is found in the limit $\beta\rightarrow0$, where the normal gradient of the charge density diverges 
as the charge density distribution $\rho(z)$ collapses onto the boundary in the form of a surface charge density, as discussed in Appendix~\ref{App:LinHydro}. In general, therefore, the nonlocal material response leads to a weaker interaction, which can be understood in terms of an effectively larger particle-surface separation due to the reconfiguration of charge carriers near the surface~\cite{FeibelMan_ProgSurfSci_12_287_1982,Reyes_PRA_72_034102_2005}, as illustrated in the inset.

\subsubsection{Convergence}
\label{Eq:nonlocal_convergence}
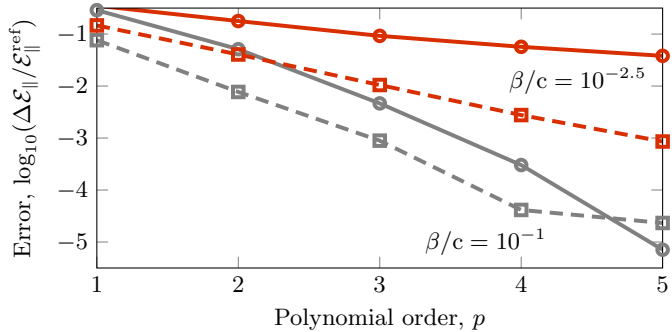
\begin{figure}
\flushright
%
%
\definecolor{mycolor1}{rgb}{0.50196,0.50196,0.50196}%
\definecolor{mycolor2}{rgb}{0.84706,0.18039,0.00000}%
\begin{tikzpicture}
\draw[draw=none, use as bounding box](0,0) rectangle (0.87\columnwidth,0.4\columnwidth);
\begin{axis}[%
compat=1.3, 
width=.87\columnwidth,
height=0.4\columnwidth,
scale only axis,
axis on top,
xmin=1,
xmax=5,
xtick={1, 2, 3, 4, 5, 6, 7},
ymin=-5.5,
ymax=-0.5,
ytick={-5, -4, -3, -2, -1},
xlabel style={align=center},
xlabel={Polynomial order, $p$},
ylabel style={align=center},
ylabel={Error, $\log_{10}(\Delta\mathcal{E}_{\|}/\mathcal{E}_{\|}^\text{ref})$},
axis background/.style={fill=white}
]
\node at (axis cs:4.4,-1.9){$\beta/\text{c}=10^{-2.5}$};
\node at (axis cs:3.75,-5){$\beta/\text{c}=10^{-1}$};
\addplot [color=mycolor1, line width=1.5pt, mark=o, mark options={solid, mycolor1}, forget plot]
  table[row sep=crcr]{%
1	-0.540039797102847\\
2	-1.29076467133748\\
3	-2.33364710215135\\
4	-3.51951492930741\\
5	-5.14829801927289\\
};
\addplot [color=mycolor2, line width=1.5pt, mark=o, mark options={solid, mycolor2}, forget plot]
  table[row sep=crcr]{%
1	-0.45473241909007\\
2	-0.749262951692314\\
3	-1.0343487654325\\
4	-1.24526127670871\\
5	-1.41915642693685\\
};
\addplot [color=mycolor1, dash pattern={on 5pt off 3pt}, line width=1.5pt, mark=square, mark options={solid, mycolor1}, forget plot]
  table[row sep=crcr]{%
1	-1.11770214950499\\
2	-2.11329131006765\\
3	-3.05019114326432\\
4	-4.38313973416225\\
5	-4.63379526618693\\
};
\addplot [color=mycolor2, dash pattern={on 5pt off 3pt}, line width=1.5pt, mark=square, mark options={solid, mycolor2}, forget plot]
  table[row sep=crcr]{%
1	-0.833746672633704\\
2	-1.39295615401888\\
3	-1.97979556065077\\
4	-2.55803995296424\\
5	-3.06655634803088\\
};
\end{axis}
\end{tikzpicture}%
\\[7mm]
\caption{\label{Fig:CasPolder_FIGold_100nm_varBeta_z_20nm_log10error_vs_order}Average relative error in the calculations in Fig.~\ref{Fig:CasimirPolderAnalytical_FIGold_100nm_z_20nm_Es_vs_b} as a function of basis function polynomial order $p$. Red and gray curves correspond to $\beta/\text{c}=10^{-2.5}$ and $\beta/\text{c}=10^{-1}$. Solid lines with circular markers were calculated using a standard mesh similar to that in Fig.~\ref{Fig:cylEmbedded_Hcyl_1p5_Hair_4p0_Pad_0p5_R_4p0_d1_0p2_h_0p1_v2}, and the dashed lines with square markers were calculated by extreme local refinement below the surface in the center of the mesh to properly sample the charge density.
} \end{figure}

Given the typical size of $\lambda_{\rm TF}$, a description of the electromagnetic response near the interface can lead to difficulties for many numerical calculation schemes, because of the need to resolve the extreme gradients in the charge density. In the present case, the problem is visible in Fig.~\ref{Fig:CasimirPolderAnalytical_FIGold_100nm_z_20nm_Es_vs_b} as a deviation between the reference curve and the numerical values indicated by red circles. By locally changing the mesh to introduce a very fine sampling close to the surface, it is possible to resolve the extreme charge density gradient and recover the reference values, as shown by the black squares.
The numerical calculations corresponding to the red circles in Fig.~\ref{Fig:CasimirPolderAnalytical_FIGold_100nm_z_20nm_Es_vs_b} were all done on the same mesh similar to that in Fig.~\ref{Fig:cylEmbedded_Hcyl_1p5_Hair_4p0_Pad_0p5_R_4p0_d1_0p2_h_0p1_v2} but with $R=300$ nm. 
To assess the convergence in detail, Fig.~\ref{Fig:CasPolder_FIGold_100nm_varBeta_z_20nm_log10error_vs_order} shows the average relative error in the numerical results as a function of basis function polynomial order and for two different values of $\beta$ corresponding to the extremes in Fig.~\ref{Fig:CasimirPolderAnalytical_FIGold_100nm_z_20nm_Es_vs_b}. The average was taken over seven nominally identical positions close to the center. The difference in exponential convergence rates signifies the increasing difficulty in resolving the charge density distribution resulting from increasingly smaller values of $\beta$. For sufficiently large polynomial order, the relative error is expected to reach a floor set by reflections from the calculation domain boundary as in Fig.~\ref{Fig:CasPolder_FIGold_100nm_z_20nm_log10error_vs_order}. 
For relatively small values of $\beta$, the steepness of the charge density and the resulting slow convergence make it difficult to reach a satisfactory error level with this particular mesh. 
To properly resolve the charge density in the case of small values of $\beta$, we can change the mesh locally by introducing small elements very close to the interface. In this way we can effectively change the convergence rate, as shown by the dashed curves in Fig.~\ref{Fig:CasPolder_FIGold_100nm_varBeta_z_20nm_log10error_vs_order}. 

\section{Repulsive Casimir-Polder force with a nonlocal material response}
\label{Sec:Applications}
\begin{figure}[htb]
\centering
\begin{overpic}[width=\columnwidth]{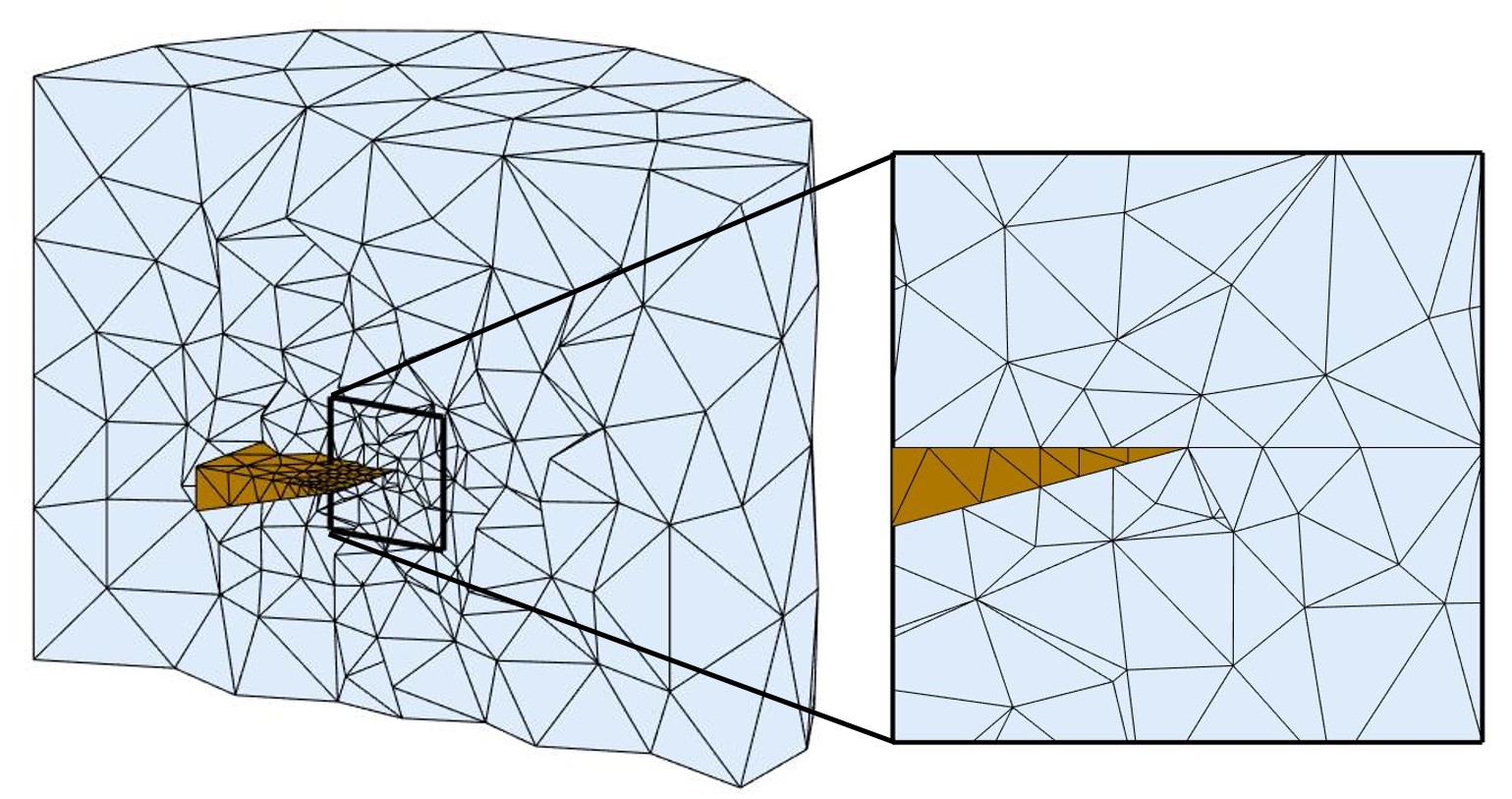}

\end{overpic}
\caption{\label{Fig:fig11.jpg}Calculation mesh for the investigation of repulsive Casimir-Polder forces near a metallic wedge. The zoom-in on the right shows a side view of the mesh in the square surface used for the Casimir-Polder energy map.} 
\end{figure}

As a non-trivial example of the flexibility of the calculation scheme, we now consider the Casimir-Polder force on a particle with anisotropic polarizability close to a sharp wedge, as illustrated in Fig.~\ref{Fig:prism_sketch}. As noted in the introduction, previous work has investigated a similar setup for a perfectly conducting wedge~\cite{Lukosz73,Milton_JPhysA_45_374006_2012}, and this setup is known for giving rise to a repulsive force along the direction orthogonal to the plane of the wedge. In this example, we revisit the wedge and examine how the use of the local Drude model or the nonlocal hydrodynamic model for the permittivity influences the repulsive component of the Casimir-Polder force. Figure~\ref{Fig:fig11.jpg} shows the calculation mesh consisting of a wedge embedded in a background of vacuum. The wedge is cut with an opening angle of $15^\circ$ from a block with cross section $100\,\text{nm}\times100\,\text{nm}$. The smallest side length of the unstructured mesh is $h\approx2$\,nm, and we note that the tetrahedra conform perfectly to the mathematically sharp edge of the wedge, so that we expect no numerical error associated with the representation of the geometry. Building on the experience from Sections~\ref{Sec:Casimir_Polder_force} and \ref{Sec:Non_local_material_response}, we consider both the case of a local Drude model in Eq.~(\ref{Eq:Epsilon_Drude}) and compare to calculations with a nonlocal permittivity as described by the hydrodynamic model in Eq.~(\ref{hydroM}).

Physically, the particle polarizability can describe the response of a microscopic anisotropic body~\cite{Levin_PRL_105_090403_2010,Rodriguez13b} or the dominant transition in an anisotropic molecule. In the following, we consider  
the molecule dibenzoterrylene, which has recently attracted attention in the literature in connection with nanophotonic systems \cite{Sadeq2018,Polisseni16,Wang17a,Gurlek21}.  
In particular, we consider the ground state to $S_1$-state transition with the associated transition dipole moment oriented along the main axis of the molecule, which we take to be parallel to the $z$-axis. Further, we substantially simplify the description by setting $\alpha_i(\omega)=0$ for $i=x,\,y$ and $\alpha_{z}(\omega) = \alpha(\omega)$, where $\alpha(\omega)$ has the same form as in Eq.~\eqref{Eq:alpha_ho} but with $\omega_\text{a}=1.58\,\text{eV}$ and $\gamma_\text{a}=7.3 \cdot 10^{-8}\,\text{eV}$ \cite{Sadeq2018}. 
\begin{figure}
\centering
\input{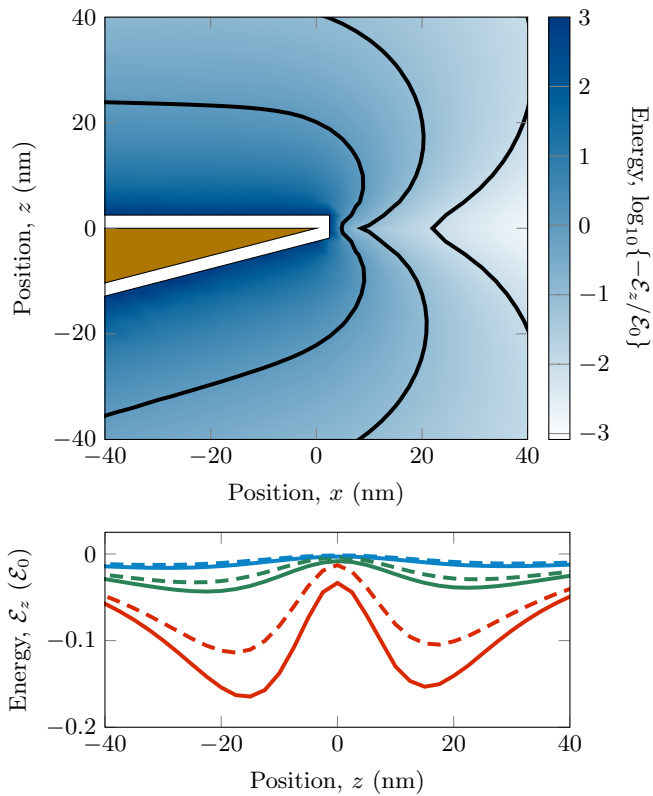}
\\[12mm]
%
%
\definecolor{mycolor1}{rgb}{0.84706,0.16078,0.00000}%
\definecolor{mycolor2}{rgb}{0.16863,0.50588,0.33725}%
\definecolor{mycolor3}{rgb}{0.04314,0.51765,0.78039}%
\begin{tikzpicture}
\draw[draw=none, use as bounding box](0,0) rectangle (0.715\columnwidth,0.3\columnwidth);
\begin{axis}[%
compat=1.3, 
width=0.715\columnwidth,
height=0.3\columnwidth,
scale only axis,
axis on top,
xmin=-40,
xmax=40,
xtick={-40, -20,   0,  20,  40},
ymin=-0.2,
ymax=0.025,
ytick={-0.2, -0.1,    0},
xlabel style={align=center},
xlabel={Position, $z$ (nm)},
ylabel style={align=center},
ylabel shift = -1pt,
ylabel={Energy, $\mathcal{E}_{z}$ ($\mathcal{E}_0$)},
axis background/.style={fill=white}
]
\addplot [color=mycolor1, line width=1.5pt, forget plot]
  table[row sep=crcr]{%
-50	-0.0327724641673851\\
-47.5	-0.0376568605699093\\
-45	-0.0433237098166802\\
-42.5	-0.0498810994406938\\
-40	-0.0574360492563305\\
-37.5	-0.0661305043325287\\
-35	-0.0760273603787777\\
-32.5	-0.0872444061621817\\
-30	-0.0997244183466636\\
-27.5	-0.113300023403747\\
-25	-0.127594871944209\\
-22.5	-0.141622850269793\\
-20	-0.15403145553743\\
-17.5	-0.162910304739634\\
-15	-0.164567386540267\\
-12.5	-0.157064027532792\\
-10	-0.137669931281841\\
-7.5	-0.107995156173333\\
-5	-0.0732954099618015\\
-2.5	-0.0441110228132462\\
0	-0.0334412865090013\\
2.5	-0.044117413763442\\
5	-0.0706643486653635\\
7.5	-0.100452029395545\\
10	-0.13004460315966\\
12.5	-0.146681477997733\\
15	-0.153202235615269\\
17.5	-0.150363909142917\\
20	-0.140665822506059\\
22.5	-0.128832055367644\\
25	-0.114883986580559\\
27.5	-0.101201883384738\\
30	-0.0883499306072455\\
32.5	-0.0766486060624439\\
35	-0.0662377421098568\\
37.5	-0.0571308807200979\\
40	-0.0492290913960431\\
42.5	-0.042461051854873\\
45	-0.0366234451350694\\
47.5	-0.0316329171708005\\
50	-0.02736313177791\\
};
\addplot [color=mycolor2, line width=1.5pt, forget plot]
  table[row sep=crcr]{%
-50	-0.0194639758525363\\
-47.5	-0.0216066349660766\\
-45	-0.0239262374449861\\
-42.5	-0.0264128354258251\\
-40	-0.0290354201651769\\
-37.5	-0.031758597513569\\
-35	-0.0345001372584511\\
-32.5	-0.0371491440527721\\
-30	-0.0395822564901432\\
-27.5	-0.0415838476297002\\
-25	-0.0429336201228338\\
-22.5	-0.0433822815331338\\
-20	-0.0426758144666645\\
-17.5	-0.0405433248336683\\
-15	-0.0368991125392254\\
-12.5	-0.0318495555607292\\
-10	-0.0258904414806605\\
-7.5	-0.0195713368466329\\
-5	-0.0139354737197312\\
-2.5	-0.00990109586324605\\
0	-0.00828217923404951\\
2.5	-0.00927370423689979\\
5	-0.0126496823755028\\
7.5	-0.0176848560374626\\
10	-0.0234194537033073\\
12.5	-0.0288630258625804\\
15	-0.0333769429919562\\
17.5	-0.0366826258506323\\
20	-0.0385782760672489\\
22.5	-0.0390547755428289\\
25	-0.0384845419010215\\
27.5	-0.0371041015400798\\
30	-0.0351508979268328\\
32.5	-0.0328339709769369\\
35	-0.03032670591594\\
37.5	-0.0277650169791959\\
40	-0.0252500910712991\\
42.5	-0.0228532773894607\\
45	-0.0205962975360475\\
47.5	-0.0185078288300828\\
50	-0.0165917001037205\\
};
\addplot [color=mycolor3, line width=1.5pt, forget plot]
  table[row sep=crcr]{%
-50	-0.0110510698924394\\
-47.5	-0.0118738996379819\\
-45	-0.0126979968177397\\
-42.5	-0.0135017649198633\\
-40	-0.0142581138452017\\
-37.5	-0.0149388376580546\\
-35	-0.0155015093694488\\
-32.5	-0.0159032756823825\\
-30	-0.0160986054275331\\
-27.5	-0.0160420239944024\\
-25	-0.0156922237593557\\
-22.5	-0.0150020778513938\\
-20	-0.0139860125887801\\
-17.5	-0.0126539362305105\\
-15	-0.0110538868245215\\
-12.5	-0.00928436000962615\\
-10	-0.00748114374504982\\
-7.5	-0.0058025803264914\\
-5	-0.00441155774605749\\
-2.5	-0.00345576222500498\\
0	-0.00303515174963532\\
2.5	-0.00318947832148825\\
5	-0.00388581745905449\\
7.5	-0.00502648411479386\\
10	-0.00646757418147195\\
12.5	-0.00805227496161534\\
15	-0.00962647734779034\\
17.5	-0.0110511711738203\\
20	-0.0122458055571338\\
22.5	-0.0131552781740043\\
25	-0.0137513984225251\\
27.5	-0.0140489857167432\\
30	-0.0140792371332191\\
32.5	-0.0138824150857335\\
35	-0.0135006949877977\\
37.5	-0.012980428674375\\
40	-0.0123499266890485\\
42.5	-0.0116579657730955\\
45	-0.0109288211884911\\
47.5	-0.010187134376688\\
50	-0.0094494815146161\\
};
\addplot [color=mycolor1, dash pattern={on 5pt off 3pt}, line width=1.5pt, forget plot]
  table[row sep=crcr]{%
-50	-0.0284241696563202\\
-47.5	-0.0324832645927949\\
-45	-0.037130448264815\\
-42.5	-0.042432929471049\\
-40	-0.0484374828667016\\
-37.5	-0.055203596831728\\
-35	-0.0627127359280526\\
-32.5	-0.0709682932616177\\
-30	-0.0798002862550991\\
-27.5	-0.088925944881933\\
-25	-0.0978916882881515\\
-22.5	-0.105836671750978\\
-20	-0.111624761593508\\
-17.5	-0.113875347908339\\
-15	-0.110377841640284\\
-12.5	-0.100386130971031\\
-10	-0.083071979202764\\
-7.5	-0.0606131539794397\\
-5	-0.0369424943785389\\
-2.5	-0.0191840353476034\\
0	-0.0128624693154424\\
2.5	-0.0203883493427069\\
5	-0.0384988841941218\\
7.5	-0.0604420067572343\\
10	-0.0812795847833464\\
12.5	-0.0956458899139503\\
15	-0.103320180697757\\
17.5	-0.104615594137058\\
20	-0.100846240803724\\
22.5	-0.0946686921589929\\
25	-0.0864156194796903\\
27.5	-0.0776708778852363\\
30	-0.069003612588532\\
32.5	-0.0607877588617002\\
35	-0.0532410904575232\\
37.5	-0.0464644142085522\\
40	-0.0404461182225117\\
42.5	-0.0351902324283963\\
45	-0.0305852721542232\\
47.5	-0.0265941492830698\\
50	-0.0231384350569431\\
};
\addplot [color=mycolor2, dash pattern={on 5pt off 3pt}, line width=1.5pt, forget plot]
  table[row sep=crcr]{%
-50	-0.0165900298204182\\
-47.5	-0.0183056643094692\\
-45	-0.0201334753002969\\
-42.5	-0.0220554411786827\\
-40	-0.0240372695346208\\
-37.5	-0.0260356496652188\\
-35	-0.0279725246281628\\
-32.5	-0.0297505881744145\\
-30	-0.0312634139207732\\
-27.5	-0.0323409595108746\\
-25	-0.0328175360772866\\
-22.5	-0.0325220918205202\\
-20	-0.0313016185821833\\
-17.5	-0.0290220444431562\\
-15	-0.0256899139260719\\
-12.5	-0.0214646470081301\\
-10	-0.0167389555565544\\
-7.5	-0.0119906934482025\\
-5	-0.00791384169383381\\
-2.5	-0.00514238760627577\\
0	-0.00411942477619665\\
2.5	-0.00500554514700693\\
5	-0.00758288537054442\\
7.5	-0.01136025858959\\
10	-0.015700755879851\\
12.5	-0.0199583331687503\\
15	-0.0236330675486854\\
17.5	-0.0264910738466394\\
20	-0.0283308502409333\\
22.5	-0.0291687461550522\\
25	-0.0291784440691533\\
27.5	-0.0285242993843076\\
30	-0.0273687975122785\\
32.5	-0.0258645885452086\\
35	-0.0241451080349713\\
37.5	-0.0223206277429137\\
40	-0.0204780632475856\\
42.5	-0.01868291450038\\
45	-0.0169572752177026\\
47.5	-0.0153364347062681\\
50	-0.0138283567369432\\
};
\addplot [color=mycolor3, dash pattern={on 5pt off 3pt}, line width=1.5pt, forget plot]
  table[row sep=crcr]{%
-50	-0.00930321245000427\\
-47.5	-0.00993878720119591\\
-45	-0.0105614911381738\\
-42.5	-0.0111520316029565\\
-40	-0.0116869564344996\\
-37.5	-0.0121424686418534\\
-35	-0.0124847377491778\\
-32.5	-0.0126805215600999\\
-30	-0.0126961498456227\\
-27.5	-0.0124999664662761\\
-25	-0.0120656905024493\\
-22.5	-0.0113691657469976\\
-20	-0.010426811533631\\
-17.5	-0.00926014603528285\\
-15	-0.00791516406998471\\
-12.5	-0.00647482107860393\\
-10	-0.00504536741661943\\
-7.5	-0.00374581993226531\\
-5	-0.00269362671285915\\
-2.5	-0.00199025758311162\\
0	-0.0017057105064808\\
2.5	-0.00186023973126438\\
5	-0.00242612163028163\\
7.5	-0.00333208186160843\\
10	-0.00447405719148631\\
12.5	-0.00573626264490849\\
15	-0.00700569394490563\\
17.5	-0.00817805192039307\\
20	-0.00918853289174098\\
22.5	-0.00999103118779369\\
25	-0.0105582009719719\\
27.5	-0.0108952895936515\\
30	-0.0110207466437907\\
32.5	-0.0109611620071873\\
35	-0.0107461598375252\\
37.5	-0.0104091779258859\\
40	-0.00997343558748015\\
42.5	-0.00947598800290174\\
45	-0.00893692490540949\\
47.5	-0.00837689817757121\\
50	-0.00781038164078073\\
};
\end{axis}
\end{tikzpicture}%
\\[7mm]
\caption{\label{Fig:prismEmbedded_Hprism_0p5_Hair_1p0_R_0p5_h_0p05_meshPlot}
Top: Contour plot showing the Casimir-Polder energy landscape in the $xy$-plane through the center of the wedge. The contour lines are exponentially spaced corresponding to the values $\log_{10}\{-\mathcal{E}_z/\mathcal{E}_0\}=-2,\,-1,\,0$.
Bottom: Calculated Casimir-Polder energy in units of $\mathcal{E}_0$ along lines parallel to the $z$-axis at $x=20\,$nm (red), $x=30\,$nm (green), and $x=40\,$nm (blue). Solid and dashed lines show results using local and nonlocal ($\beta/\text{c}=10^{-1.5}$) material responses, respectively.
}
\end{figure}

The top panel of Fig.~\ref{Fig:prismEmbedded_Hprism_0p5_Hair_1p0_R_0p5_h_0p05_meshPlot}
shows a contour plot of the calculated Casimir-Polder energy as a function of position in the $xz$-plane through the center of the wedge at $y=0$ for the case of the nonlocal hydrodynamic model. The cross section of the wedge is superimposed on the picture to mark its position. It is surrounded by a white spacing to block off the immediate vicinity of the material surface, where additional physical phenomena are expected to influence the particle. In this limit, the model must ultimately be augmented with additional terms to account for effects such as tunneling and charge-transfer as well as the granularity and atomic structure of the material. The bending of the contour lines towards the tip of the wedge signifies the interesting region of a repulsive component of the Casimir-Polder force. To appreciate this effect more clearly, the bottom panel of Fig.~\ref{Fig:prismEmbedded_Hprism_0p5_Hair_1p0_R_0p5_h_0p05_meshPlot} shows the calculated energy along three lines in the plane $y=0$, parallel to the $z$-axis and for different values of $x$, corresponding to different distances to the wedge. The solid lines show the Casimir-Polder energy for the case of a local material description, while the dashed lines indicate the nonlocal response in Eq.~\eqref{hydroM} with $\beta/\text{c}=10^{-1.5}$. 
In all curves, local minima along the $z$-direction are clearly visible on either side of the wedge with deeper minima closer to the tip. In the full three-dimensional map of the energy, however, these minima correspond to saddle points preventing a stable equilibrium in accordance with the results of Ref.~\cite{Rahi_PRL_070404_2010}. These curves were calculated using polynomal basis functions of order $p=4$. To estimate the accuracy, we compare to the same curves calculated with $p=3$ (not shown) and find a maximum relative deviation of $\text{max}\{\Delta\mathcal{E}_z/\mathcal{E}_z\}\approx0.03$. Combined with the convergence studies in Fig.~\ref{Fig:CasPolder_FIGold_100nm_z_20nm_log10error_vs_order}, for which the reflections from the boundary are expected to be larger than in the present case, we consider this relative deviation to be a very conservative estimate of the maximum error in the calculations. 

\begin{figure}
\flushright
%
%
\definecolor{mycolor1}{rgb}{0.84706,0.16078,0.00000}%
\definecolor{mycolor2}{rgb}{0.16863,0.50588,0.33725}%
\definecolor{mycolor3}{rgb}{0.04314,0.51765,0.78039}%
\begin{tikzpicture}
\draw[draw=none, use as bounding box](0,0) rectangle (0.87\columnwidth,0.3\columnwidth);
\begin{axis}[%
compat=1.3, 
width=.87\columnwidth,
height=0.3\columnwidth,
scale only axis,
axis on top,
xmin=-40,
xmax=40,
xtick={-40, -20,   0,  20,  40},
ymin=-1.5,
ymax=1.5,
ytick={-1,  0,  1},
xlabel style={align=center},
xlabel={Position, $z$ (nm)},
ylabel style={align=center},
ylabel={Force, $F_z$ ($F_{0}$)},
axis background/.style={fill=white}
]
\addplot [color=mycolor1, line width=1.5pt, forget plot]
  table[row sep=crcr]{%
-50	0.195375856100968\\
-47.5	0.226673969870839\\
-45	0.262295584960542\\
-42.5	0.302197992625471\\
-40	0.347778203047928\\
-37.5	0.395874241849956\\
-35	0.448681831336161\\
-32.5	0.499200487379278\\
-30	0.543024202283335\\
-27.5	0.571793941618481\\
-25	0.561119133023357\\
-22.5	0.49634421070548\\
-20	0.355153968088162\\
-17.5	0.066283272025317\\
-15	-0.300134360298978\\
-12.5	-0.775763850038072\\
-10	-1.18699100434031\\
-7.5	-1.38798984846126\\
-5	-1.16737548594221\\
-2.5	-0.426789452169796\\
0	0.427045090177627\\
2.5	1.06187739607686\\
5	1.19150722920725\\
7.5	1.18370295056461\\
10	0.665474993522941\\
12.5	0.260830304701419\\
15	-0.113533058894059\\
17.5	-0.387923465474342\\
20	-0.473350685536605\\
22.5	-0.557922751483397\\
25	-0.547284127832818\\
27.5	-0.514078111099716\\
30	-0.468052981792063\\
32.5	-0.416434558103486\\
35	-0.364274455590354\\
37.5	-0.316071572962193\\
40	-0.270721581646806\\
42.5	-0.233504268792141\\
45	-0.199621118570759\\
47.5	-0.170791415715617\\
};
\addplot [color=mycolor2, line width=1.5pt, forget plot]
  table[row sep=crcr]{%
-50	0.0857063645416099\\
-47.5	0.0927840991563788\\
-45	0.0994639192335628\\
-42.5	0.104903389574073\\
-40	0.108927093935682\\
-37.5	0.109661589795284\\
-35	0.105960271772839\\
-32.5	0.0973244974948466\\
-30	0.0800636455822782\\
-27.5	0.0539908997253432\\
-25	0.0179464564120005\\
-22.5	-0.0282586826587713\\
-20	-0.0852995853198489\\
-17.5	-0.145768491777714\\
-15	-0.201982279139848\\
-12.5	-0.238364563202749\\
-10	-0.252764185361103\\
-7.5	-0.225434525076068\\
-5	-0.161375114259407\\
-2.5	-0.0647566651678617\\
0	0.0396610001140115\\
2.5	0.135039125544121\\
5	0.201406946478392\\
7.5	0.229383906633787\\
10	0.217742886370924\\
12.5	0.180556685175032\\
15	0.132227314347044\\
17.5	0.0758260086646638\\
20	0.0190599790232018\\
22.5	-0.0228093456722952\\
25	-0.0552176144376689\\
27.5	-0.0781281445298819\\
30	-0.0926770779958343\\
32.5	-0.100290602439877\\
35	-0.102467557469763\\
37.5	-0.100597036315873\\
40	-0.0958725472735342\\
42.5	-0.0902791941365272\\
45	-0.083538748238591\\
47.5	-0.0766451490544893\\
};
\addplot [color=mycolor3, line width=1.5pt, forget plot]
  table[row sep=crcr]{%
-50	0.0329131898217008\\
-47.5	0.0329638871903147\\
-45	0.0321507240849415\\
-42.5	0.030253957013535\\
-40	0.0272289525141191\\
-37.5	0.0225068684557666\\
-35	0.0160706525173489\\
-32.5	0.00781318980602492\\
-30	-0.00226325732522973\\
-27.5	-0.0139920094018672\\
-25	-0.0276058363184785\\
-22.5	-0.0406426105045464\\
-20	-0.0532830543307819\\
-17.5	-0.0640019762395625\\
-15	-0.0707810725958137\\
-12.5	-0.0721286505830528\\
-10	-0.0671425367423368\\
-7.5	-0.0556409032173564\\
-5	-0.0382318208421006\\
-2.5	-0.0168244190147865\\
0	0.00617306287411723\\
2.5	0.0278535655026498\\
5	0.0456266662295748\\
7.5	0.0576436026671236\\
10	0.0633880312057356\\
12.5	0.0629680954470002\\
15	0.056987753041199\\
17.5	0.0477853753325409\\
20	0.0363789046748183\\
22.5	0.0238448099408331\\
25	0.011903491768724\\
27.5	0.0012100566590334\\
30	-0.00787288189942074\\
32.5	-0.0152688039174342\\
35	-0.020810652536908\\
37.5	-0.0252200794130588\\
40	-0.0276784366381215\\
42.5	-0.0291657833841757\\
45	-0.0296674724721218\\
47.5	-0.0295061144828778\\
};
\addplot [color=mycolor1, dash pattern={on 5pt off 3pt}, line width=1.5pt, forget plot]
  table[row sep=crcr]{%
-50	0.162363797458989\\
-47.5	0.185887346880805\\
-45	0.212099248249359\\
-42.5	0.240182135826106\\
-40	0.270644558601055\\
-37.5	0.300365563852983\\
-35	0.330222293342605\\
-32.5	0.353279719739254\\
-30	0.365026345073359\\
-27.5	0.358629736248737\\
-25	0.317799338513053\\
-22.5	0.231523593701191\\
-20	0.0900234525932721\\
-17.5	-0.139900250722221\\
-15	-0.399668426770116\\
-12.5	-0.692566070730678\\
-10	-0.898353008932972\\
-7.5	-0.946826384036032\\
-5	-0.71033836123742\\
-2.5	-0.252862641286438\\
0	0.301035201090579\\
2.5	0.724421394056596\\
5	0.877724902524503\\
7.5	0.833503121044482\\
10	0.574652205224156\\
12.5	0.306971631352261\\
15	0.051816537572043\\
17.5	-0.150774133333359\\
20	-0.247101945789243\\
22.5	-0.330122907172103\\
25	-0.349789663778159\\
27.5	-0.346690611868173\\
30	-0.328634149073271\\
32.5	-0.301866736167082\\
35	-0.27106704995884\\
37.5	-0.240731839441617\\
40	-0.210235431764618\\
42.5	-0.184198410966925\\
45	-0.159644914846134\\
47.5	-0.138228569045069\\
};
\addplot [color=mycolor2, dash pattern={on 5pt off 3pt}, line width=1.5pt, forget plot]
  table[row sep=crcr]{%
-50	0.0686253795620418\\
-47.5	0.0731124396331059\\
-45	0.0768786351354347\\
-42.5	0.0792731342375215\\
-40	0.0799352052239229\\
-37.5	0.0774749985177599\\
-35	0.0711225418500672\\
-32.5	0.0605130298543471\\
-30	0.0431018236040548\\
-27.5	0.0190630626564825\\
-25	-0.0118177702706568\\
-22.5	-0.0488189295334742\\
-20	-0.0911829655610852\\
-17.5	-0.133285220683373\\
-15	-0.169010676717673\\
-12.5	-0.189027658063026\\
-10	-0.189930484334079\\
-7.5	-0.163074070174746\\
-5	-0.110858163502322\\
-2.5	-0.0409185132031649\\
0	0.0354448148324113\\
2.5	0.1030936089415\\
5	0.151094928761825\\
7.5	0.17361989161044\\
10	0.170303091555968\\
12.5	0.146989375197407\\
15	0.114320251918157\\
17.5	0.0735910557717581\\
20	0.0335158365647557\\
22.5	0.000387916564043716\\
25	-0.026165787393827\\
27.5	-0.0462200748811655\\
30	-0.060168358682795\\
32.5	-0.0687792204094922\\
35	-0.072979211682305\\
37.5	-0.0737025798131218\\
40	-0.0718059498882263\\
42.5	-0.0690255713070955\\
45	-0.0648336204573814\\
47.5	-0.0603231187729928\\
};
\addplot [color=mycolor3, dash pattern={on 5pt off 3pt}, line width=1.5pt, forget plot]
  table[row sep=crcr]{%
-50	0.0254229900476659\\
-47.5	0.0249081574791138\\
-45	0.0236216185913093\\
-42.5	0.0213969932617227\\
-40	0.0182204882941552\\
-37.5	0.0136907642929743\\
-35	0.00783135243688323\\
-32.5	0.000625131420912002\\
-30	-0.0078473351738611\\
-27.5	-0.0173710385530754\\
-25	-0.0278609902180667\\
-22.5	-0.0376941685346647\\
-20	-0.0466666199339251\\
-17.5	-0.0537992786119257\\
-15	-0.0576137196552312\\
-12.5	-0.05717814647938\\
-10	-0.0519818993741647\\
-7.5	-0.0420877287762465\\
-5	-0.0281347651899011\\
-2.5	-0.0113818830652331\\
0	0.00618116899134357\\
2.5	0.0226352759606897\\
5	0.0362384092530722\\
7.5	0.0456790131951152\\
10	0.050488218136887\\
12.5	0.0507772519998855\\
15	0.0468943190194977\\
17.5	0.0404192388539164\\
20	0.0320999318421085\\
22.5	0.0226867913671276\\
25	0.0134835448671863\\
27.5	0.00501828200556478\\
30	-0.00238338546413496\\
32.5	-0.00860008678648514\\
35	-0.0134792764655691\\
37.5	-0.0174296935362311\\
40	-0.0198979033831366\\
42.5	-0.0215625238996901\\
45	-0.0224010691135309\\
47.5	-0.0226606614716194\\
};
\end{axis}
\end{tikzpicture}%
\\[7mm]
\caption{\label{Fig:Fz_vs_zPos_20_30_40nm_local_and_beta_1em3}Calculated Casimir-Polder forces corresponding to the energies in the bottom panel of Fig.~\ref{Fig:prismEmbedded_Hprism_0p5_Hair_1p0_R_0p5_h_0p05_meshPlot}. The expressions are normalized with respect to $F_{0}=\mathcal{E}_{0}/L_{0}$. At all three distances, the local description (solid lines) leads to a stronger modulation of the force than the nonlocal material response (dashed lines).
}
\end{figure}
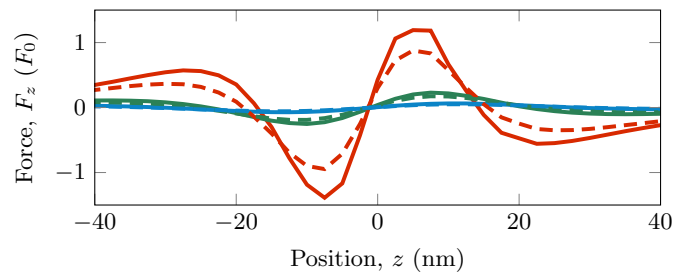

Although the Casimir-Polder energy is seen to be clearly affected by the introduction of a nonlocal material response, the effect is predominantly to shift the energy levels and to smoothen the curves, resulting in somewhat shallower minima. As a consequence, the curves look qualitatively the same, although there is an associated reduction in the force. 
In order to see this more clearly, 
Fig.~\ref{Fig:Fz_vs_zPos_20_30_40nm_local_and_beta_1em3} shows the $z$-component of the Casimir-Polder force $F_\text{z}(\mr)=-\partial_z\mathcal{E}_z(\mr)$ along the same three lines as in Fig.~\ref{Fig:prismEmbedded_Hprism_0p5_Hair_1p0_R_0p5_h_0p05_meshPlot}.  
The influence of the nonlocal material response is especially visible for the line closest to the wedge. 
In all cases the largest magnitude of the force becomes smaller for a nonlocal material response as a result of the smearing out of the charge density close to the sharp edge. 
For visualization purposes, and to reduce the computational requirements, in these calculations we used a value of $\beta$ which is larger than what can be found in realistic materials. A reduction of the value of $\beta$ will bring the curves closer to those corresponding to the local Drude model. From Figs.~\ref{Fig:prismEmbedded_Hprism_0p5_Hair_1p0_R_0p5_h_0p05_meshPlot} and \ref{Fig:Fz_vs_zPos_20_30_40nm_local_and_beta_1em3} we conclude therefore, that the repulsive Casimir-Polder force component near the sharp edge is robust towards the modifications introduced by the nonlocal material response.

\section{Conclusion}
\label{Sec:Conclusion}

Over the last decades, fluctuation-induced forces have attracted growing attention for their relevance
in fundamental sciences and for the role they can play in many nanophotonic and quantum technological devices. Accurate evaluation of these forces in non-trivial geometries is therefore becoming important. 
In this work, we have described a numerical method to perform high-accuracy calculations of the Casimir-Polder force on polarizable particles in arbitrary material structures using advanced material models. The method builds on the time-domain approach of Refs.~\cite{Rodriguez_PRA_76_032106_2007, Rodriguez_PRA_80_012115_2009, McCauley_PRA_81_012119_2009}, which we have modified in several ways. In essence, it combines a scattered-field formulation of Maxwell's equations with a non-trivial post-processing of the numerical signal in time by means of appropriately designed and analytically known kernel functions. These modifications enable an easy implementation in existing time-domain solvers, in particular the Discontinuous Galerkin Time-Domain (DGTD) method, which we have used in this work. Since changes in the particle polarizability tensor affect only the kernel function, variations in the polarizability can be easily implemented in post-processing of the same set of simulation data. 
The use of DGTD offers a number of interesting possibilities in terms of unstructured meshing, exponential convergence, and moderate memory usage. We have presented a detailed discussion of the convergence properties in configurations involving local and nonlocal material models. Specifically, we have considered the Drude model and the linearized hydrodynamic model, respectively, and we have argued that the accuracy is limited primarily by numerical discretization and reflections from the calculation domain boundary. By increasing the polynomial order of the basis functions, we recovered the expected exponential convergence, and owing to the notoriously fast reduction of Casimir-type forces with distance, we have illustrated how the reflections from the boundaries can be very effectively controlled by increasing the calculation domain. Indeed, comparing to independent reference calculations, we have found relative errors as low as a few parts in a million. 
As a practical application example, we have demonstrated the flexibility of the method by investigating the Casimir-Polder interaction between an anisotropic particle and a sharp wedge in vacuum. Similar structures with sharp edges have been investigated before and are known to give rise to a repulsive component of the Casimir-Polder force. Comparing the results using a Drude model to that of a linearized hydrodynamic model, we found the repulsive force prediction from the local response model to be remarkably robust against introduction of nonlocal material responses even when using parameters for the non-locality that are much larger than in realistic materials. 

In closing, we remark on a number of different possibilities for refinement and generalization of the calculation method. For the practical calculations, one can likely improve the performance by locally varying the polynomial order of the basis functions. For the calculations with nonlocal materials, in particular, this will provide an interesting alternative to locally refining the mesh in order to accurately resolve charge density gradients near the interfaces. Similarly, as already discussed in the main text, the method as presented considers only the case of zero temperature, but the effect of a finite temperature can be implemented by a corresponding modification of the kernel function. More complex geometries and more involved material models~\cite{Horovitz12, Reiche19, Hannemann21, Mortensen21a, Pfeifer22, Wegner23} can be considered to further investigate how the interplay of these factors impacts the behavior of the Casimir-Polder force. Conversely, an accurate comparison between measurements and theoretical predictions can provide information about the physics of such material systems. Also we note, that if the particle features a magnetic response, there will be an additional magnetic Casimir-Polder interaction, which can be evaluated using a modification of the scheme presented in this article. See also Refs. \cite{Haakh09b,Reiche20} for a discussion about the magnetic Casimir-Polder interaction and its interplay with nonlocal material responses. Last, we note that since the method in essence follows the ideas of Refs.~\cite{Rodriguez_PRA_76_032106_2007, Rodriguez_PRA_80_012115_2009, McCauley_PRA_81_012119_2009}, it can be naturally extended in order to exploit the advantages of the DGTD method for evaluation of the Casimir force between macroscopic bodies. Moreover, following previous work~\cite{Rodriguez2011}, we expect that the scheme can be generalized to nonequilibrium configurations and thereby enable calculations of a wider spectrum of fluctuation-induced phenomena \cite{Volokitin03a,Volokitin07,Chapuis08,Reiche17,Reiche19}. 

\section{Acknowledgements}
\label{Sec:Acknowledgements}
We thank Daniel Reiche, Marty Oelschl{\"a}ger, Julia Werra, and Dan-Nha Huynh for valuable help and discussions. P.T.K. acknowledges funding from the Carlsberg foundation and the Danish National Research Foundation through NanoPhoton - Center for Nanophotonics (DNRF147).

\appendix
\section{Scattered field formulation and material models}
\label{App:ScatteredFieldFormulation}

The calculations presented in the main text rely on the numerical evaluation of the electromagnetic field from a point dipole. 
This field, in turn, can be conveniently calculated by separating the total electromagnetic field in two parts corresponding to the incoming and the scattered part of the field, respectively, as
\begin{align}
\mE^\text{T}(\mr,t) &= \mE^\text{I}(\mr,t) + \mE^\text{S}(\mr,t)\label{Eq:ET_def}\\
\mH^\text{T}(\mr,t) &= \mH^\text{I}(\mr,t) + \mH^\text{S}(\mr,t).
\end{align}
By linearity of the Maxwell equations, it follows that the scattered part of the fields fullfill equations similar to the original Maxwell curl equations, but with an artificial source term governed by the incoming field $\mE^\text{I}(\mr,t)$ \cite{Busch_LaserPhotRev_5_773_2011}. This reformulation alleviates the need for an explicit numerical simulation of the source by recasting the Maxwell equations in a form that accounts only for the scattering due to materials embedded in an otherwise homogeneous and non-dispersive medium, which we take to be vacuum in this work. The incoming part of the field is taken to be the field from a point dipole, which is known analytically~\cite{Jackson}. In practice, the exact form of the resulting equations depends on the particular material model. Below, we list the explicit formulations used for the calculations in this work. 
For completeness, and as an illustrative starting point, we also include the artificial case of dispersionless media.

\subsection{Dispersionless media}
In dispersionless media, the scattered fields obey the equations
\begin{align}
\label{Eq:Scattered_fields_dispersionless_media}
\epsilon_0\epsilon_\text{R}(\mathbf{r})\partial_t \mE^\text{S}(\mr,t) &= \nabla\times\mH^\text{S}(\mr,t) -\epsilon_0\Delta\epsilon(\mathbf{r})\partial_t\mE^\text{I}(\mr,t) \\
\mu_0\mu_\text{R}(\mathbf{r})\partial_t \mH^\text{S}(\mr,t) &= -\nabla\times\mE^\text{S}(\mr,t) - \mu_0\Delta\mu(\mathbf{r})\partial_t\mH^\text{I}(\mr,t),
\end{align}
where $\Delta\epsilon(\mathbf{r})=\epsilon_\text{R}(\mathbf{r})-\epsilon_\text{B}$ and $\Delta\mu(\mathbf{r})=\mu_\text{R}(\mathbf{r})-\mu_\text{B}$ represent the changes in relative permittivity and permeability defining the material in the otherwise homogeneous background defined by $\epsilon_\text{B}$ and $\mu_\text{B}$. In order to ensure a numerically sound implementation, we require the electric and magnetic fields at the start of the simulation to vanish. Since the field of a point dipole with a given current time dependence $J(t)$ already depends on the derivative of $J(t)$~\cite{Jackson}, the appearance of the time derivative of the incoming fields in Eq.~\eqref{Eq:Scattered_fields_dispersionless_media} puts additional constraints on the source current at $t=0$. For this reason, the source in Eq.~\eqref{Eq:Source_type_1} has continuous first and second derivatives, which both vanish at $t=0$.

\subsection{Drude model}
\label{App:DrudeModel}
Due to their widespread use in many experimental setups, it is interesting to consider material models for metals, such as the Drude model, for which the scattered fields obey the equations
\begin{align}
\label{Eq:Scattered_fields_Drude}
\epsilon_0\partial_t \mE^\text{S}(\mr,t) &= \nabla\times\mH^\text{S}(\mr,t) -\mJ(\mr,t), \\
\mu_0\partial_t \mH^\text{S}(\mr,t) &= -\nabla\times\mE^\text{S}(\mr,t),
\end{align}
where $\mJ(\mr,t)$ is the total current density. In the case of a local Drude model corresponding to a frequency-dependent relative permittivity of the form in Eq.~(\ref{Eq:Epsilon_Drude}), the current density is governed by the auxiliary differential equation~\cite{Wolff_OE_21_12022_2013}
\begin{align}
\partial_t\mJ(\mr,t) = \omega_\text{pl}^2\mE^\text{T}(\mr,t)- \Gamma\mJ(\mr,t)~,
\label{Eq:Drude_J}
\end{align}
where now the incoming field $\mE^\text{I}(\mr,t)$, which is included in $\mE^\text{T}(\mr,t)$ in accordance with Eq.~(\ref{Eq:ET_def}), acts as an additional driving term for the current density for the purpose of calculating the scattered fields. In the Drude model, the current density is directly proportional to the electric field in the frequency domain, so we do not need to specify additional boundary conditions for $\mJ(\mr,t)$ at the boundaries of the metal.  

\subsection{Linearized hydrodynamics}
\label{App:LinHydro}
To investigate the impact of materials with nonlocal optical responses on Casimir-Polder forces, we consider the linearized hydrodynamic model~\cite{Moeferdt18}, for which the longitudinal part of the dielectric function is given by Eq.~\eqref{hydroM}. In this model, the current density is coupled to a charge density $\rho(\mr,t)$ and the two fields are governed by the auxiliary differential equations
\begin{align}
\partial_t \rho(\mr,t) &= -\nabla\cdot\mathbf{J}(\mr,t) 
\label{Eq_HD_dt_rho} \\
\partial_t \mathbf{J}(\mr,t) &= \omega_\text{pl}^2\mE^\text{T}(\mr,t) - \Gamma\mathbf{J}(\mr,t) - \beta^2\nabla\rho(\mr,t),
\label{Eq:hydrodynamic_J}
\end{align}
where the parameter $\beta$ is the speed of sound, which describes the propagation of a density perturbation within the electronic fluid. While Eq.~(\ref{Eq_HD_dt_rho}) is simply a continuity equation, Eq.~(\ref{Eq:hydrodynamic_J}) can be seen as a simplified version of more accurate descriptions~\cite{Mortensen21a}. 
In the case of a Fermi fluid, it follows from these more elaborate models that the parameter $\beta$ scales linearly with the Fermi velocity~\cite{Moeferdt18,Reiche19,Wegner23}. We note, that since the charge density couples only to the longitudinal part of the current density, the transverse part of the equation reduces to Eq.~(\ref{Eq:Drude_J}). As a consequence, both the transverse and the longitudinal part of the dielectric function are correctly described by Eqs.~(\ref{Eq_HD_dt_rho}) and (\ref{Eq:hydrodynamic_J}).

The hydrodynamic equations must be augmented by additional boundary conditions (ABCs) at the metal-vacuum interfaces, which must be inferred from physical arguments~\cite{Baryakhtar91,Baryakhtar91a}. Assuming specular reflection of the electrons at the interface~\cite{Kliewer68,Barton79,Ford84,Schmidt16,Reiche20,Mortensen21a}, the normal component of the current density is required to vanish at the boundary.  
Since, in principle, the tangential component may be finite this ABC is also known as the slip boundary condition. This approach implicitly assumes that the material-vacuum interface can be treated as a sharp boundary and that the electron dynamics are those of the bulk material at all positions inside but arbitrarily close to the surface. We note that this is an idealization, and that more advanced descriptions of the interface itself have been developed to treat the electromagnetic field at surfaces in detail~\cite{FeibelMan_ProgSurfSci_12_287_1982,Mortensen21a}.   
If the root-mean-square roughness of the surface becomes comparable to the mean free path of the bulk electrons, for example, 
the conditions for specular reflection might no longer be applicable and diffusive scattering processes or other approaches should be taken into account \cite{Kliewer70,Kaganova01,Wegner23}.

Due to the pressure term in Eq.~\eqref{Eq:hydrodynamic_J}, there can be no surface charge density for finite values of $\beta$. As a result, we must require the normal component of the electric field to be continuous across the interface, which is very different from the condition that one would find in the case of a local material model. To investigate this limit, consider the equation for the normal component of the current density at the boundary. Since $J_\perp(\mathbf{r},t)=0$, it follows that $\partial_tJ_\perp(\mathbf{r},t)=0$, and just inside the boundary we find
\begin{align}
\nabla_\perp \rho(\mr,t) = \frac{1}{\lambda^{2}} E_\perp(\mr,t),
\label{Eq:non_local_BC_gradient_rho}
\end{align}
where $\perp$ denotes the vector component in the direction normal to the surface, and $\lambda=\beta/\omega_{\rm pl}$ describes the range of variation of the charge density within the material and close to the interface. Comparing the low-frequency limit of Eq.~(\ref{hydroM}) to the Thomas-Fermi dielectric function~\cite{Ashcroft76}
\begin{align}
\epsilon_{\rm TF}(k)= 1 + \frac{1}{\lambda_\text{TF}^2k^2},
\label{Eq:epsilon_TF}
\end{align}
we find that $\lambda$ is exactly the Thomas-Fermi screening length~\cite{Reiche17,Reiche19}.  
Given a finite electric field component $E_\perp(\mr,t)$, Eq.~\eqref{Eq:non_local_BC_gradient_rho} results in a sharp rise in the charge density near the boundary. For numerical methods, this requires a very fine sampling of the field near the surface in order to resolve the large gradient. 

In the limit $\beta=0$, Eqs.~\eqref{Eq_HD_dt_rho} and \eqref{Eq:hydrodynamic_J} decouple and the model reduces to the local Drude model. In this case, the denominator in Eq.~\eqref{Eq:non_local_BC_gradient_rho} vanishes, which is consistent with the collapse of the volume charge density onto the boundary as a surface charge density. 

\section{Kernel function}
\label{App:Calculation_of_g}
The function $g_{i}(t)$ is defined from its Fourier transform in Eq.~\eqref{Eq:g_of_omega} which, in turn, depends on the Fourier transform of the source current variation $J(t)$. Given that the source is nonzero only for $t>0$, we can define the frequency dependence of the source formally as $J(\omega)=\lim_{\xi\to 0^+}J(\zeta)$, where
\begin{equation}
J(\zeta) = \int_{0}^{\infty} J(t)\text{e}^{\mathrm{i}\zeta t}\ud t, \quad \zeta=\omega+\mathrm{i}\xi,
\label{Eq:J_def_integral}
\end{equation}
with $\xi> 0$. For all choices of $J(t)$ for which it exists, this limit defines $J(\omega)$ as the analytical continuation onto the real frequency axis of the function $J(\zeta)$ in the upper half of the complex-frequency plane. This definition enables the use of a large ensemble of source current variations, including in principle an undamped sinusoidal behavior.
Since $J(t)$ is real and causal, it follows that $J(\zeta)=J^{*}(-\zeta^{*})$ and that $J(\zeta)$ is analytic in the upper half of the complex-frequency plane.  The same properties are true for the functions $\alpha_i(\omega)$. Integrating Eq.~(\ref{Eq:J_def_integral}) by parts, we find that
\begin{equation}
J(\zeta)\sim \frac{J^{(n)}(0)}{(-\mathrm{i}\zeta)^{n+1}}, |\zeta|\to \infty, 
\end{equation}
where $J^{(n)}(0)$ indicates the first nonzero $n$'th derivative evaluated at $t=0$. 
For large values of $|\zeta|$, the function $1/J(\zeta)$ behaves as $1/J(\zeta)\sim \left(-\mathrm{i}\zeta\right)^{n+1}$. Since $J(\zeta)$ is analytic for $\xi>0$, in the limit $\zeta \to 0$ it has a Taylor expansion, where the derivative has to be evaluated in the limit of $\xi=0^{+}$. 
Having defined and analyzed $J(\zeta)$, we now turn to the properties of $g_{i}(t)$. Following the definition in Eq.~\eqref{Eq:g_of_omega} and after performing a Wick rotation in the first quadrant of the complex-frequency plane, we can write that
\begin{align}
\label{Generag}
\mathrm{Im}[g_{i}(-t)]
=&\frac{1}{2\pi}\int_{0}^{\infty}\frac{\xi \alpha_i(\mathrm{i}\xi)}{J(\mathrm{i}\xi)}\text{e}^{-\xi t}\ud \xi
\nonumber\\
&+\mathrm{Im}\left[\sideset{}{'}\sum_{k} \text{e}^{\mathrm{i}\zeta_{k} t}\mathrm{Res}\left[\frac{\zeta\alpha_i(\zeta)}{J(\zeta)}\right]_{\zeta=\zeta_{k}}\right],
\end{align}
where $\zeta_{k}\not=0$ are the zeros of $J(\zeta)$ in the upper half of the complex plane, and for simplicity we have assumed that they all have multiplicity one. The prime in the sum corresponds to a prefactor $1/2$ for real or purely imaginary zeros. Whereas the sum in Eq.~\eqref{Generag} evidently leads to oscillations, which are exponentially damped in general, the temporal behavior of the integral depends on details of the current variation. Nevertheless, we can investigate the limiting behaviors for general current variations as follows. For large times, integration by parts shows that 
\begin{equation}
\label{longt}
\int_{0}^{\infty}\frac{\xi \alpha_i(\mathrm{i}\xi)}{J(\mathrm{i}\xi)}\text{e}^{-\xi t}\ud \xi\sim-\frac{\left[\frac{\xi \alpha_i(\mathrm{i}\xi)}{J(\mathrm{i}\xi)}\right]^{(l)}_{\vert\xi\to 0^{+}}}{(-t)^{l+1}}, t\to \infty, 
\end{equation}
where the symbol $\left[\cdots\right]^{(l)}_{\vert\xi\to 0^{+}}$ indicates the first nonzero $l$'th derivative of the function in parentheses evaluated in the limit $\xi\to 0^{+}$. For small times, on the other hand, 
the relevant behavior of the integral is determined by the behavior of the integrand in the limit $\xi \to \infty$. 
Assuming that $\alpha_i(\mathrm{i}\xi)\sim \alpha_0/ \xi^{m}$ for $\xi \to \infty$ and defining $y=\xi t$, we find that 
\begin{align}
\int_{0}^{\infty} \frac{\xi \alpha_i(\mathrm{i}\xi)}{J(\mathrm{i}\xi)} \text{e}^{-\xi t}\ud \xi
&=\frac{1}{t^{2}}\int_{0}^{\infty}\frac{y \alpha_i(\mathrm{i}y/t)}{J(\mathrm{i}y/t)}\text{e}^{-y}\ud y\\
&\sim\frac{(n+2-m)!}{t^{n+3-m}} 
\frac{ \alpha_0}{J^{(n)}(0)}, t\to 0.
\label{shortt}
\end{align}

\subsection{Source-specific behavior}

In this work, we use the source current variation given in Eq.~(\ref{Eq:Source_type_1}), for which the first nonzero derivative of $J(t)$ at $t=0$ is  
$J^{(3)}(0)= 24\gamma^{3}J_{0}$, 
and for which the Fourier transform $J(\omega)$ is given by
\begin{equation}
\label{Jomega}
J(\omega)
=
- 24 \text{i} J_0 \gamma^3 
\frac{\omega}{(\gamma - \text{i} \omega)^5}.
\end{equation}
Since $J(\omega)$ vanishes only for $\omega=0$, the temporal behavior of the kernel function is determined entirely by the first term 
on the right-hand-side of 
Eq.~\eqref{Generag}. The behavior at large times is therefore given by Eq.~(\ref{longt}), and assuming $\alpha_i(0)\neq 0$, we find that
\begin{equation}
\lim_{\xi\to 0^{+}}\frac{\alpha_i(\xi)\xi}{J(\xi)}=\frac{\alpha_i(0) \gamma^2}{24J_0},
\end{equation}
from which we can identify $l=0$ as the first nonzero derivative. For any nonvanishing static polarizability, therefore, the kernel function always goes to zero as $\mathcal{O}(t^{-1})$ in the limit $t\to \infty$. Moreover, from the form of the polarizability in Eq.~\eqref{Eq:alpha_ho}, for which $\alpha_i(\mathrm{i}\xi)\sim\alpha_{0}\omega_\text{a}^{2}\xi^{-2}$ for $\xi \to \infty$, we identify $m=2$, and by Eq.~(\ref{shortt}) it follows that $\mathrm{Im}[g_i(-t)]$ diverges as $\mathcal{O}(t^{-4})$ for $t\to 0$. 

In the artificial case of a constant polarizability of the form $\alpha_i(\omega)=\alpha_0$, we identify $m=0$, so that $\mathrm{Im}[g_{i}(-t)]$ diverges as $\mathcal{O}(t^{-6})$ for $t\to 0$. In this case, the simplicity of the expression in Eq.~\eqref{Jomega} enables a direct analytical calculation of the kernel function, for which we find
\begin{align}
\mathrm{Im}[g_i(-t)] 
&=\frac{\alpha_{0}}{2\pi}\int_{0}^{\infty}\text{e}^{-\xi t}   
\frac {(\gamma + \xi)^5}{24 J_0 \gamma^3} \ud \xi 
\nonumber\\
&=\frac{\alpha_{0}\gamma^{3}}{2\pi J_0}\sum_{k=0}^{5}\frac{\frac{5}{(5-k)!}}{(\gamma t)^{k+1}},
\label{Eq:g_from_Source_type_1}
\end{align} 
which is consistent with the general predictions above for the limiting behavior at long and short times. 

An analytical expression for the kernel function can be calculated also for the dispersive polarizability in Eq.~\eqref{Eq:alpha_ho}. To this end, we make use of the fact that the polarizability is analytic in the upper half of the complex plane and transform the integral to the positive imaginary frequency axis on which the polarizability takes the form
\begin{align}
\alpha(\mathrm{i}\xi) &= \frac{\alpha_0 \omega_\text{a}^2}{\left(\xi +\mathrm{i}\Omega\right)\left(\xi -\mathrm{i}\Omega^{*}\right)}
=- \mathrm{Im}\left[\frac{\alpha_0\omega_\text{a}^2}{\Omega_\text{R}\left(\xi +\mathrm{i}\Omega\right)}\right],
\end{align}
where $\Omega = \sqrt{\omega_\text{a}^2 - \gamma_\text{a}^2/4}-\mathrm{i}\gamma_\text{a}/2$, in which $\omega_\text{a}$ and $\gamma_\text{a}$ are both positive and $\Omega_\text{R}=\mathrm{Re}[\Omega]$. 
Inserting in Eq.~(\ref{Eq:g_of_omega}) and carrying out the integral, one finds that
\begin{align}
\mathrm{Im}[g(-t)]
&=-\frac{\alpha_0\omega_\text{a}^2}{48\pi J_0\Omega_\text{R}\gamma^3}\mathrm{Im}\left[\int_{0}^{\infty}\frac{ (\gamma + \xi)^5}{\xi +\mathrm{i}\Omega}\text{e}^{-\xi t}\ud \xi\right]
\nonumber\\
&=-\frac{\alpha_0\omega_\text{a}^2}{48\pi J_0\Omega_\text{R}\gamma^3}\mathrm{Im}\left[\left(\gamma-\mathrm{i}\Omega\right)^{5} \phantom{\sum^{5}_{k=0}}\right.
\nonumber\\
&\qquad\qquad\quad\times\left.\sum^{5}_{k=0}\binom{5}{k}\frac{\text{e}^{\mathrm{i}\Omega t}\Gamma[k,\mathrm{i}\Omega t]}{[\left(\gamma-\mathrm{i}\Omega\right)t]^{k}}\right],
\end{align}
where $\Gamma[s,z]$ is the upper incomplete gamma function,
\begin{equation}
\Gamma[s,z]=\int_{z}^{\infty} t^{s-1}\text{e}^{-t}\ud t.
\end{equation}
For $s\equiv k\in \mathbb{N}^{+}$ the gamma function can be written as
\begin{equation}
 \Gamma[k,z]=(k-1)!\text{e}^{-z}\sum_{j=0}^{k-1}\frac{z^{j}}{j!},
\end{equation}
so that we can rewrite $\mathrm{Im}[g(-t)]$ as
\begin{widetext}
\begin{align}
\mathrm{Im}[g(-t)]
=-\frac{\alpha_0\omega_\text{a}^2\gamma^2}{48\pi J_0\Omega_\text{R}}\mathrm{Im}\left[
\left(1-\frac{\text{i}\Omega}{\gamma}\right)^{5}\text{e}^{\text{i}\Omega t}\Gamma[0,\mathrm{i}\Omega t]
-
\sum^{5}_{j=1}
\frac{\sum_{k=j}^{5}\frac{5!\left(1-\frac{\text{i}\Omega}{\gamma}\right)^{5-k}\left(\frac{\text{i}\Omega}{\gamma}\right)^{k-j}}{k(k-j)!(5-k)!}}{(\gamma t)^{j}}
\right]
.
\end{align}
\end{widetext}
As a consistency check we can inspect the asymptotic expression. Given that
\begin{equation}
\Gamma[0,\mathrm{i}\Omega t]\sim
\begin{cases}
\frac{\text{e}^{-\mathrm{i} t \Omega }}{\mathrm{i} t \Omega }&t\to \infty\\
\mathrm{i} t \Omega -\ln [\mathrm{i} \Omega t]-\gamma_{\rm E}& t\to0,
\end{cases}
\end{equation}
where $\gamma_{\rm E}$ is the Euler–Mascheroni constant, we have that
\begin{equation}
\label{Eq:g_from_Source_type_2}
\mathrm{Im}[g(-t)]\sim
\begin{cases}
\frac{\alpha_0\gamma^3}{48\pi J_0 }\frac{1}{\gamma t}
 &t\to \infty\\
\frac{\alpha_0\omega_\text{a}^2\gamma}{8\pi J_0 }\frac{1}{(\gamma t)^{4}}& t\to0,
\end{cases}
\end{equation}
which again is consistent with the general predictions for the limiting behavior at long and short times.

\end{document}